\documentclass{aastex631}

  \newcommand{\noise}{{\mathcal N}}

  \usepackage{amsmath}
  \usepackage{graphicx}
  \usepackage[utf8]{inputenc}
\renewcommand{\arraystretch}{1.5}
  \graphicspath{{./}{figures/}}

    \defcitealias{hanasoge19}{HM19}
\defcitealias{mandal2020}{MH20}
\defcitealias{mandal21}{MHG21}
  \shorttitle{Inertial Waves in the Sun}
\begin{document}
\title{Probing Depth Variations of Solar Inertial Modes through Normal Mode Coupling}

   \author[0000-0003-3067-288X]{Krishnendu Mandal}
  \affiliation{New Jersey Institute of Technology, Newark, NJ 07102, USA}
  
  \author[0000-0003-2896-1471]{Shravan M. Hanasoge}
  \correspondingauthor{Krishnendu Mandal}
  \email{krishnendu.mandal@njit.edu}
  \affiliation{Department of Astronomy \& Astrophysics,
  Tata Institute of Fundamental Research, Mumbai 400005}
  \affiliation{Center for Space Science, New York University Abu Dhabi, PO Box 129188, Abu Dhabi, UAE}
  
   \begin{abstract}
    Recently discovered inertial waves, observed on the solar surface, likely extend to the deeper layers of the Sun. Utilizing helioseismic techniques, we explore these motions, allowing us to discern inertial-mode eigenfunctions in both radial and latitudinal orientations. We analyze $8$ years of space-based observations ($2010 - 2017$) taken by the Helioseismic and Magnetic Imager (HMI) onboard the Solar dynamic observatory (SDO) using normal-mode coupling. Coupling between same and different-degree acoustic modes and different frequency bins are measured in order to capture the various length scales of inertial modes. We detect inertial modes at high latitude with azimuthal order $t=1$ and frequency $\sim -80$ nHz. This mode is present in the entire convection zone. The presence of Rossby modes may be seen down to a depth of $\sim 0.83R_\odot$ and the Rossby signal is indistinguishable from noise below that depth for high azimuthal order. We find that the amplitudes of these modes increase with depth down to around $0.92 R_\odot$ and decrease below that depth. We find that the latitudinal eigenfunctions of Rossby modes deviate from sectoral spherical harmonics if we use a similar approach as adopted in earlier studies. We found that spatial leakage and even pure noise in the measurements of non-sectoral components can also explain the above-mentioned characteristics of the latitudinal eigenfunctions. This realization underscores the necessity for careful interpretation when considering the latitudinal eigenfunctions of Rossby modes. Exploring the depth-dependent characteristics of these modes will enable us to capture interior dynamics distinctly, separate from p-mode seismology. 
   
  \end{abstract}
  
  \section{Introduction} \label{sec:intro}
 The Sun supports a large number of inertial modes, e.g., equatorial Rossby modes, high-latitude inertial modes and critical latitudinal modes \citep{gizon21}. Previously, \citet{gizon18} detected equatorial Rossby modes \citep[first observed in earth by][]{rossby_1939} in the Sun in which Coriolis force is the restoring mechanism \citep{Papa1978,saio_82,provost_81}. Since Coriolis force is the primary driver of these modes, their frequency is also of the same order as the rotation frequency of the Sun. 
 \subsection{Equatorial Rossby modes}
Equatorial Rossby modes have a well-defined dispersion relation based on their harmonic degree, $s$ and azimuthal order, $t$,  
\begin{flalign}
\sigma &=-\frac{2t\Omega}{s(s+1)}\nonumber\\ 
           & =-\frac{2\Omega}{(t+1)} \hspace{0.2cm} \textrm{when }\hspace{0.2cm} s=t ,
\label{eq:dispersionRossby}
\end{flalign} 
in a co-rotating frame. In this study, we opt for tracking at the surface equatorial rotation rate of $\Omega/(2\pi)=453.1$ nHz. We utilize harmonic degree and azimuthal order, denoted as $(\ell, m)$, to characterize solar acoustic modes, and employ $(s, t)$ to represent inertial waves in subsequent sections. Observed Rossby modes closely follow this dispersion relation. Various methods of helioseismology have been used to confirm this finding, i.e.,  mode coupling \citep[][hereafter HM19, MH20, MHG21 repectively]{hanasoge19,mandal2020,mandal21}, ring-diagram analysis \citep{hill88,proxauf2020,hanson20}, and time-distance helioseismology \citep{duvall, liang_2018}. Although these prior studies were able to characterize the modes at the surface, the depth dependence has not been determined. Ring-diagram analysis and local correlation tracking are only sensitive to the near-surface layers. \citet{proxauf2020} attempted to estimate the depth variations of Rossby modes from the surface down to 8 Mm using ring-diagram analysis, finding that amplitudes decrease with depth. \citet{hathaway2020}, using supergranulation tracking, found mode amplitudes to be invariant with depth. Though \citet{provost_81} derived the equations for the depth dependence of these modes for a uniformly rotating medium, solving that equation is not straightforward. \citet{damiani2020} solved the equation derived by \citet{provost_81} numerically for different polytropes in the inviscid limit. They found that only sectoral modes with no radial nodes may exist and that the radial dependencies of the modes scale as $r^t$. Since the Sun is not a uniformly rotating inviscid polytrope, their results cannot be directly applied to the Sun. Moreover, careful consideration is required at the base of the convection zone, a sharp layer that separates the convectively stable radiative interior from the unstable zone above. Additionally, the Sun rotates differentially both in latitude and radius, whereas Equation $11$ of \citet{provost_81} was derived for a uniformly rotating medium. Since Rossby modes extend in both latitude and radius, differential rotation arguably \citep{dziembowski87_1,gizon2020} significantly influences these modes. Recently, \citet{bekki22_linear,triana22,jishnu23}  developed numerical methods which take into account solar stratification, turbulent diffusivities, differential rotation and a latitudinal entropy gradient to study inertial waves in the Sun. In particular, \citet{bekki22_linear} assumed spatially uniform values for viscosity, thermal diffusivity and superadiabaticity to simplify the problem, which are not true for the Sun. Additionally, their computational domain excluded the tachocline and near-surface shear layer. How important those are in determining the radial eigenfunction still needs to be determined. \citet{bekki22_linear,jishnu23} found that there exist different radial orders of Rossby modes, e.g., $n=0$ and $n=1$, which are associated with different frequencies at low azimuthal orders, but almost identical frequencies at high azimuthal order. In their study, \citet{dikpati22}, based on 3D hydrodynamic shallow-water simulations of a rotating thin spherical shell, suggested that Rossby waves within supergranular layers may be excited through the inverse cascade of kinetic energy. Exploring the depth-dependent characteristics of Rossby waves and other inertial waves becomes significant in advancing our understanding of solar interior dynamics, with far-reaching implications for predicting space-weather events.
Since our current understanding of the radial dependence of Rossby waves is still nascent, we primarily focus on inferring their behavior from SDO/HMI data and do not attempt to explain it analytical or numerical means. We also characterize the latitudinal eigenfunctions of these modes in this work. 

\subsection{High latitude inertial modes} 
Analyzing $10$ years of SDO/HMI data, \citet{gizon21} detected a swirling pattern with its maximum velocity occurring at high latitudes, concluding that it is a normal mode of the Sun. Similar features were detected earlier by \citet{hathaway13,bogart15,hathaway2020}, although they were reported as giant-cell convection at the time.  Though several analyses with different techniques have been used to confirm the existence of equatorial Rossby modes, been few studies have focused on characterizing inertial modes at high latitude, mainly because current instruments dedicated to helioseismology have limited coverage near the pole and systematical biases become significant near the limb \citep[e.g., center-to-limb systematics in time-distance helioseismology][]{zhao2012}. We apply the mode-coupling method in this work to detect this mode in the convection zone.
 
Characterizing the eigenfunctions of these inertial modes is important as they may provide information about the solar interior that is otherwise not possible to obtain by classical helioseismology. These modes are sensitive to turbulent viscosity and superadiabaticity in the convection zone \citep{bekki22_linear}. Characterizing these modes will therefore enable us to map latitudinal entropy gradient, turbulent viscosity and superadiabaticity in the convection zone. Although \citetalias{mandal2020} performed several tests that involve accurately recovering the depth profiles of Rossby modes for synthetic measurements with added noise (also see in the appendix), we additionally verify our technique by recovering the $s=3$ spatial scale of differential rotation from the surface down to the radiative interior and compare it with inferences obtained from global helioseismology.     

 \section{Data analysis} \label{sec:data}
High-resolution line-of-sight Doppler observations, $\Phi$, taken by SDO/HMI, are analyzed here. In normal-mode coupling, we transform the data fully to the spectral domain, first to the spherical-harmonic domain to obtain $\Phi_{\ell m}$, where $\ell$ and $m$ are the harmonic degree and azimuthal order of the p-mode respectively \citep[for details, see][]{larson2015}.  We then temporally Fourier transform these data to obtain $\Phi_{\ell m}^{\omega}$, where $\omega$ represents the temporal frequency. We obtain time series of these data products from the JSOC website \footnote{JSOC: \url{http://jsoc.stanford.edu/}}. As p-modes are excited by turbulent convection, we expect the cross-correlation to contain power only for the quantity $\vert \Phi^{\omega *}_{\ell m}\Phi_{\ell m}^{\omega}\vert$. For simpler notation, we express $\Phi^{\omega}_{\ell m}$ without explicitly showing its dependence on the radial order, $n$.
Cross-spectral measurements such as 
$\langle\Phi^{\omega *}_{\ell m}\Phi_{\ell+\Delta\ell m+t}^{\omega+\sigma}\rangle$ in general will not carry any information because of the stochastic nature of the p-mode excitation, unless the difference in frequency, $\sigma$, harmonic degree, $\Delta\ell$ and azimuthal order, and $t$ match with the temporal and spatial scales of perturbations in the medium. This work examines the coupling, $\langle\Phi^{\omega *}_{\ell m}\Phi_{\ell+\Delta\ell m+t}^{\omega+\sigma}\rangle$ between modes of the same radial order, $n$. A general flow field, which we consider as a perturbation to our background medium, may be expressed in terms of poloidal and toroidal components 
\begin{align}
\mathbf{u}(r,\theta,\phi;\sigma)=\sum_{s=0}^{\infty}\sum_{t=-s}^{s} u_{st}^{\sigma}(r)Y_{st}(\theta,\phi)\hat{\mathbf{r}}+v_{st}^{\sigma}(r)\nabla_1 Y_{st}(\theta,\phi)-\nonumber\\
w_{st}(r)\hat{\mathbf{r}}\times \nabla_1 Y_{st}(\theta,\phi),
\label{eq:uVel}
\end{align}
where $\theta$ and $\phi$ are co-latitude and longitude respectively, and $\nabla_1=\hat{\theta}\partial_\theta+\frac{\hat{\phi}}{\sin\theta}\partial_\phi$. The first two terms with $u^\sigma_{st}$ and $v^\sigma_{st}$ determine the poloidal component, and the final term, $w^{\sigma}_{st}$, captures toroidal motions. Mass conservation ties $u_{st}^{\sigma}$ and $v_{st}^{\sigma}$, and as a consequence, Equation \ref{eq:uVel} has only two independent components. Here, we choose $u_{st}^{\sigma}$ and $w^\sigma_{st}$ as the independent parameters. 
The cross-spectral measurement, $\langle\Phi^{\omega *}_{\ell m}\Phi_{\ell+\Delta\ell m+t}^{\omega+\sigma}\rangle$, in presence of a flow as given by Equation \ref{eq:uVel} may be expressed as \citep{hanasoge17_etal}
\begin{equation}
    \Phi^{\omega *}_{\ell m}\Phi_{\ell+\Delta\ell m+t}^{\omega+\sigma}=H^\sigma_{\ell \ell+\Delta\ell m t} \sum_{s t} 
    \int_{0}^{R_\odot} dr[u_{st}^{\sigma}(r)K^{u}_{st}(r)+w_{st}^{\sigma}(r)K^{w}_{st}(r)],
    \label{eq:kernel}
\end{equation}
 where the expression for $H^\sigma_{\ell \ell+\Delta\ell m t}$ is given by Equation $5$ of \citetalias{mandal21}.
  $K^{u}_{st}$ and $K^{w}_{st}$ are sensitivity kernels for poloidal and toroidal components respectively.  These kernels are dependent on the modes of interest $(n,\ell,m)$ and $(n,\ell+\Delta\ell,m+t)$. For the sake of simplicity in notation, we have refrained from listing all the dependencies in $K^{u}_{st}$ and $K^{w}_{st}$, which may be simplified using their asymptotic expressions \citep{vorontsov11,hanasoge18}. 
 \begin{eqnarray}
    K^{u}_{st}= \gamma^{\ell+\Delta\ell s \ell}_{tm} g_{\Delta\ell,s}\mathcal{K}_{n\ell}(r),\\
    K^{w}_{st}= \gamma^{\ell+\Delta\ell s \ell}_{tm} f_{\Delta\ell,s}\mathcal{K}_{n\ell}(r),
    \label{eq:ku}
 \end{eqnarray}
 where the expression for $\gamma^{\ell+\Delta\ell s \ell}_{tm}$ is given by Equation $4$ of \citetalias{mandal21}. The formula for $\mathcal{K}_{n\ell}(r)$ can be found in \citet{hanasoge18} in terms of the eigenfunctions of the p-modes. The factor $f$ is given by 
 \begin{align}
     f_{\Delta\ell,s}=(-1)^{(s+\Delta\ell-1)/2}\times \frac{(s-\Delta\ell)!!(s+\Delta\ell)!!}{\sqrt{(s-\Delta\ell)!(s+\Delta\ell)!}},
     \label{eq:fS}  
 \end{align}
 for odd $s+\Delta\ell$ and the expression for $g$ is
 \begin{align}
g_{\Delta\ell,s}=i(-1)^{(s+\Delta\ell)/2}\Delta\ell\times \frac{(s-\Delta\ell-1)!!(s+\Delta\ell-1)!!}{\sqrt{(s-\Delta\ell)!(s+\Delta\ell)!}},
     \label{eq:gS}
 \end{align} 
 for even $s+\Delta\ell$. Since Rossby and inertial modes at high latitude are mostly toroidal in nature, our choice of $\Delta\ell$ is such that $s+\Delta\ell$ is always odd. For toroidal flow, Equation \ref{eq:uVel} reduces to 
\begin{align}
    \mathbf{u}(r,\theta,\phi;\sigma)=\sum_{s=0}^{\infty}\sum_{t=-s}^{s} 
    w^{\sigma}_{st}(r)\hat{\mathbf{r}}\times \nabla_1 Y_{st}(\theta,\phi).
    \label{eq:uTor}
\end{align}
 Because of the selection criterion described subsequent to Equation \ref{eq:fS}, we may only choose values $\Delta\ell=0,2,\dots$ to determine odd $s$, and $\Delta\ell=1,3,\dots$ to detect even harmonic degrees. \citetalias{hanasoge19} used $\Delta\ell=0$ to determine the odd harmonic degrees of Rossby modes. Subsequently, \citetalias{mandal21} used $\Delta\ell=1,3$ to measure even harmonic degrees of Rossby modes. We neglect the poloidal component from Equation \ref{eq:kernel}, 
 \begin{equation}
     \Phi^{\omega *}_{\ell m}\Phi_{\ell+\Delta\ell m+t}^{\omega+\sigma}=H^\sigma_{\ell \ell+\Delta\ell m t} \sum_{s} 
    \left(\int_{0}^{R_\odot} dr w_{st}^{\sigma}(r)K^{w}_{st}(r)\right).
    \label{eq:kernel_toroidal}
 \end{equation}
 We define the term in the first bracket as the B-coefficient \citep{hanasoge18}
 \begin{equation}
     B^\sigma_{s\,t}(n,\ell,\ell+\Delta\ell)=f_{\Delta\ell,s}\int_{0}^{R_\odot} dr w_{st}^{\sigma}(r)\mathcal{K}_{n\ell}(r).
     \label{eq:BInv}
 \end{equation}
 We see from Equation~\ref{eq:BInv} that B coefficients depend on the modes $(n,\ell)$ and $(n,\ell+\Delta\ell)$ but not on azimuthal order, $m$ and frequency, $\omega$. Therefore, we estimate the B coefficients from Equation \ref{eq:kernel_toroidal}  \citep{woodard16} using 
 \begin{equation}
    B^\sigma_{s\,t}(n,\ell,\ell+\Delta\ell)=\frac{\sum_{m,\,\omega}\gamma^{\ell+\,\Delta\ell\,\,s\,\ell}_{t\, m}\,H^{\sigma *}_{\ell\,\ell+\,\Delta\ell\,m\,t}(\omega)\,\Phi^{\omega *}_{\ell m}\,\Phi^{\omega+\,\sigma+\,t\Omega}_{\ell+\,\Delta\ell\,\, m+t}}{\sum_{m,\omega}\vert \gamma^{\ell+\,\Delta\ell\,\,s\,\ell}_{t\, m}\,H^{\sigma *}_{\ell\,\ell+\Delta\ell\,\,m\,t}\vert^2}.\label{eq:BCoeff}
 \end{equation}
 The summation applies over all $m$ and $\omega$ to improve the signal-to-noise ratio. We ensure that the frequency interval over which the sum is carried out satisfies the criterion 
 \begin{align}
    & \left| \omega-\omega_{n\,\ell\, m} \right| \leq \Gamma_{n\,\ell}  \\
    & \textrm{or }
    \left|\omega- \omega_{n\,\,\ell+\,\Delta\ell\,\,m+t}+(\sigma+\,t\Omega)\right|\leq 
   \Gamma_{n\,\,\ell+\,\Delta\ell}.
    \label{eq:rangeSum}
 \end{align}
 \begin{figure}
 \centering
    \includegraphics[scale=0.45]{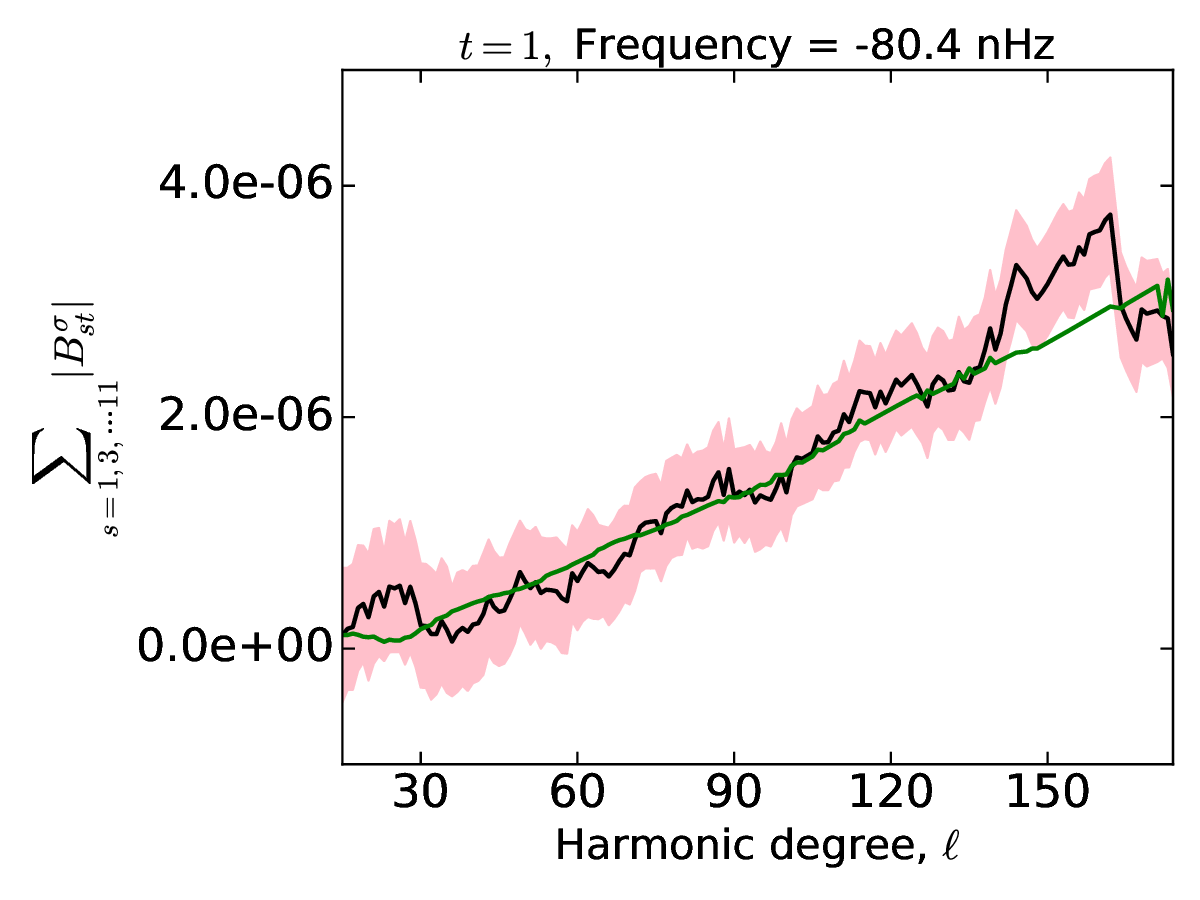}
    \caption{Measured B-coefficients (black solid lines), $B^\sigma_{st}(n,\ell)$ for the frequency bin $\sigma/(2\pi)= -80.4$ nHz, which also corresponds to the peak frequency of inertial waves at high latitude. Since B-coefficients show alternating sign with $\ell$, we multiply it by $(-1)^\ell$ and sum over all harmonic degrees, $\ell$. We subsequently average this expression over all radial orders, $n$. The corresponding errors in the measured values ($\pm 1\sigma$ around the mean) is indicated by the pink shaded area. The inverted profile, $w^\sigma_{st}$, obtained using the measured B-coefficients has been used to in forward modeling and shown using green solid line.}
    \label{fig:HL_bcoeff}
\end{figure}

  We analyze a span of $8$ years of SDO/HMI data, covering the period from $2010$ to $2017$. Our focus involves investigating p and f-modes over mode-frequency and harmonic-degree ranges of ranges $[1000, 5600]$ $\mu$Hz and $\ell\in [10, 180]$, respectively. To comprehensively examine the low-frequency evolution of toroidal flows, we focus on the frequency range $\sigma/(2\pi)\in[-500, 0]$ nHz when we track at the surface equatorial rotation rate of $\Omega/(2\pi)=453.1$ nHz.  We solve the inverse problem set out in Equation \ref{eq:BInv}, and thereby determine $w^\sigma_{st}$ through the application of a regularized least-square technique, as outlined in \citetalias{mandal2020}. By recovering $w^\sigma_{st}$ as a function of radius, we gain insights into the depth-dependent behaviour of these modes.  
 
\section{Results}    
\subsection{High-latitude mode with azimuthal order $t=1$}\label{sec:HL_mode} 
\begin{figure}
    \centering
    \includegraphics[scale=0.4]{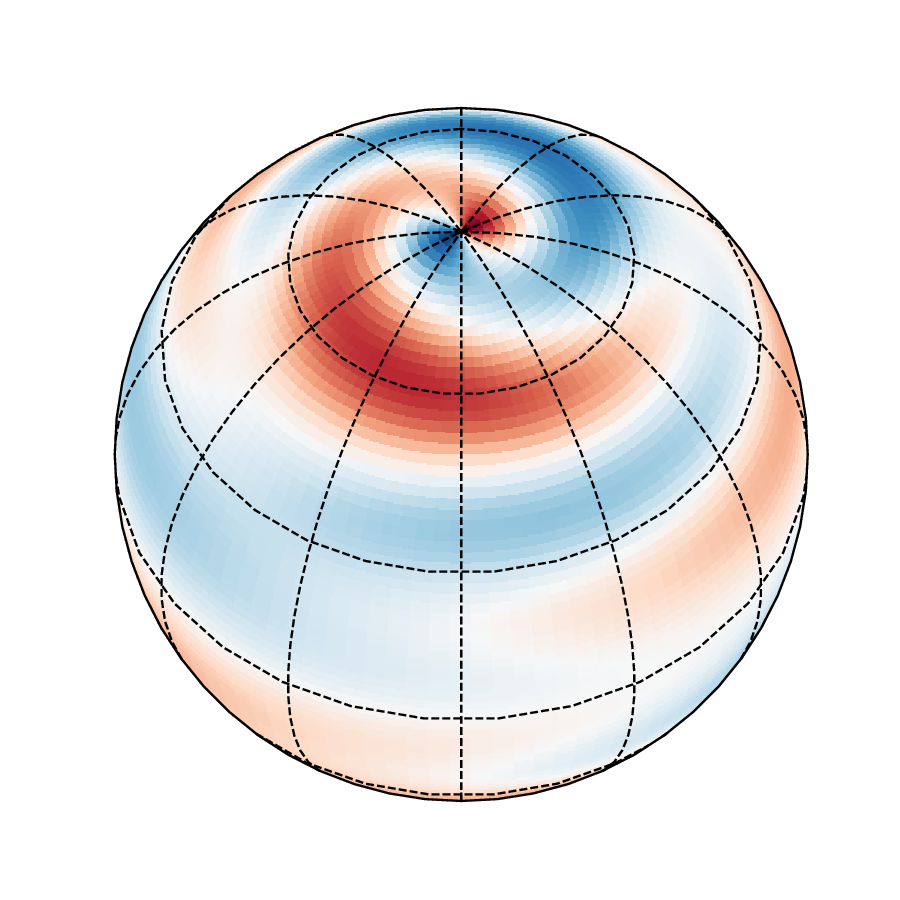}    
    \includegraphics[scale=0.4]{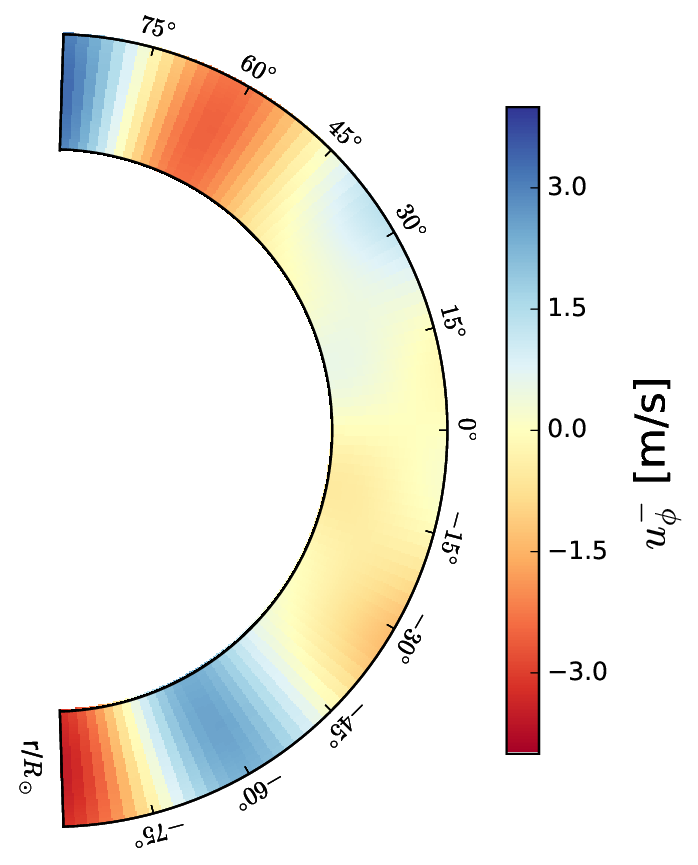}\includegraphics[scale=0.4]{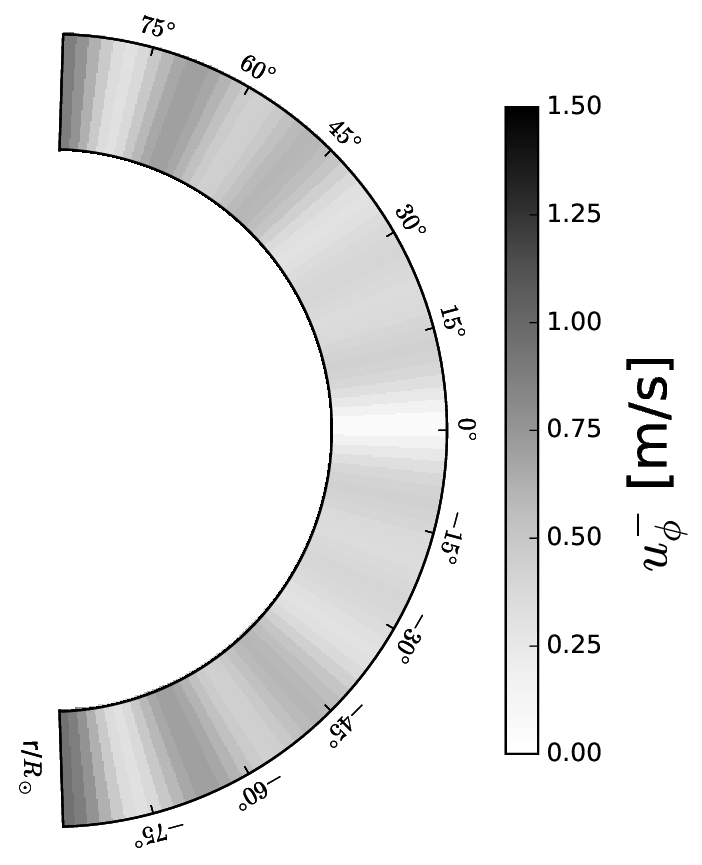}
    \includegraphics[scale=0.5]{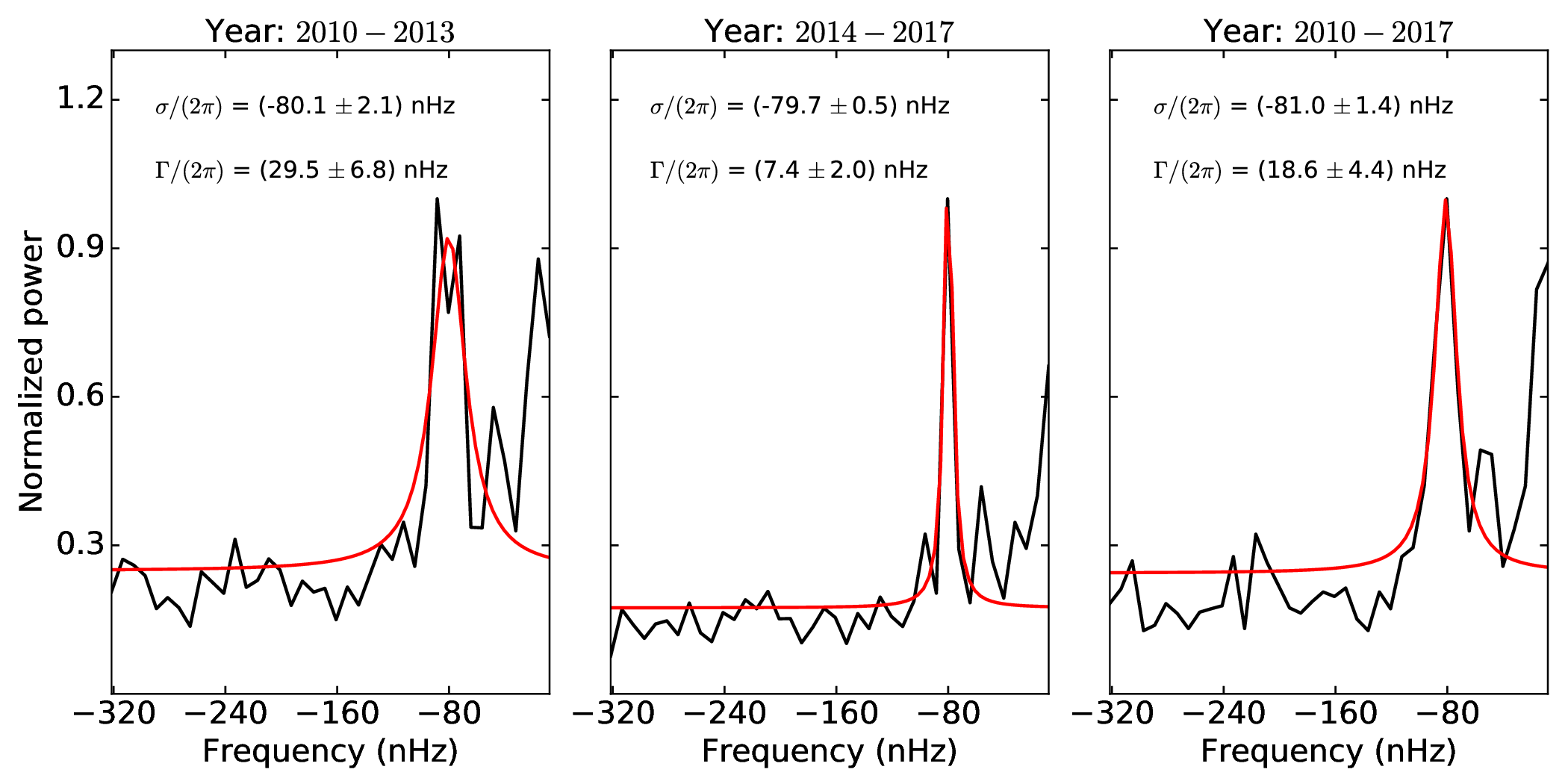}
    \caption{Upper left panel: eigenfunctions of solar inertial modes of azimuthal order $t=1$. A spiraling pattern associated with this mode is seen at high latitude. The upper middle panel displays the radial profile of $u_{\phi}^{-}$, while the corresponding error is presented in the right panel. Bottom panel: Normalized power spectra of $u_{\theta}^+$ over the entire latitude range (solid black line) for different time periods, stated in the title of each panel. The power has been normalized in the frequency versus latitude diagram for each latitude. Subsequently, we compute the mean spectrum across latitudes to generate the normalized power versus frequency diagram. We fit a Lorentzian function with a constant background to the power spectrum. The fitted values of frequency, $\sigma/(2\pi)$ and line-width, $\Gamma/(2\pi)$ are mentioned in each panel. Fits to the spectrum are shown as red solid lines.}
    \label{fig:HL_sphere}
\end{figure}

\begin{figure}
   \centering
  \includegraphics[scale=0.4]{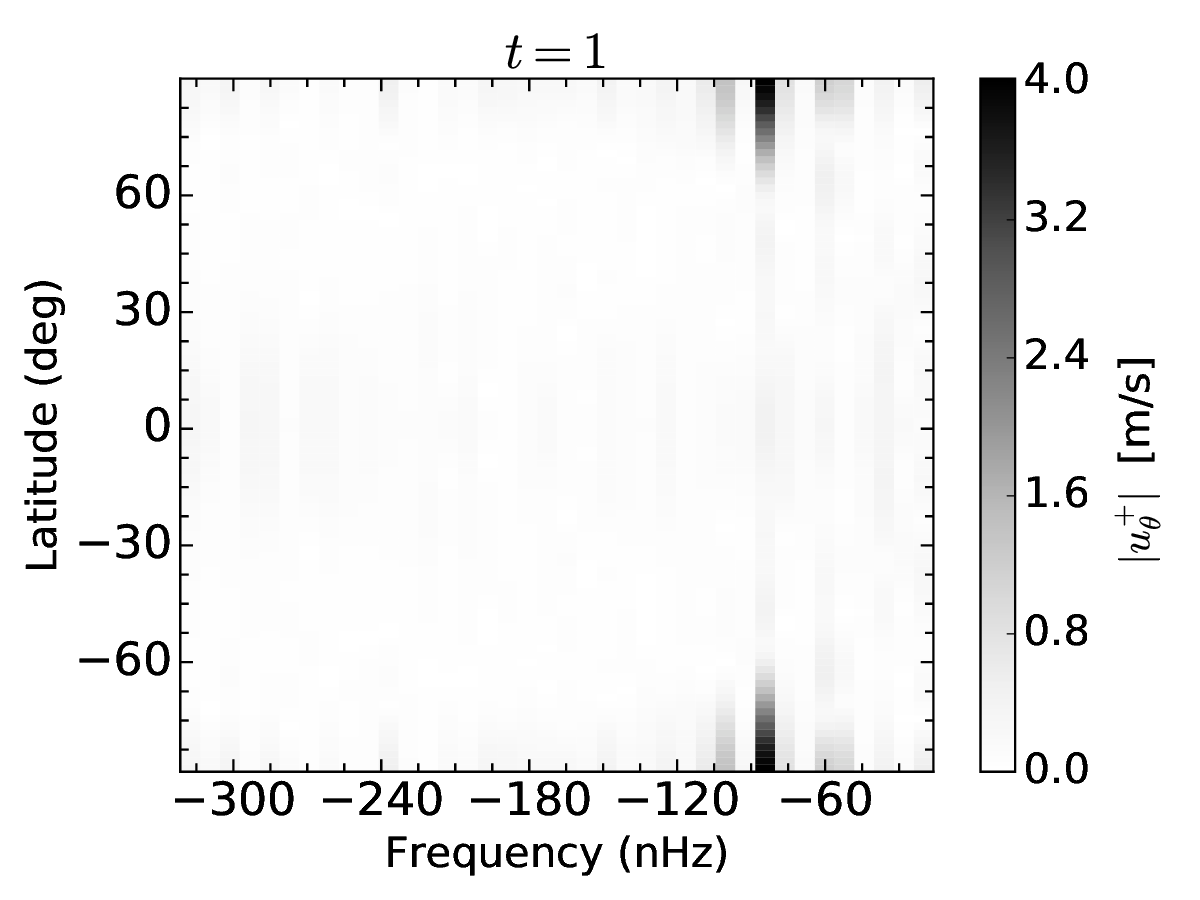}  \includegraphics[scale=0.4]{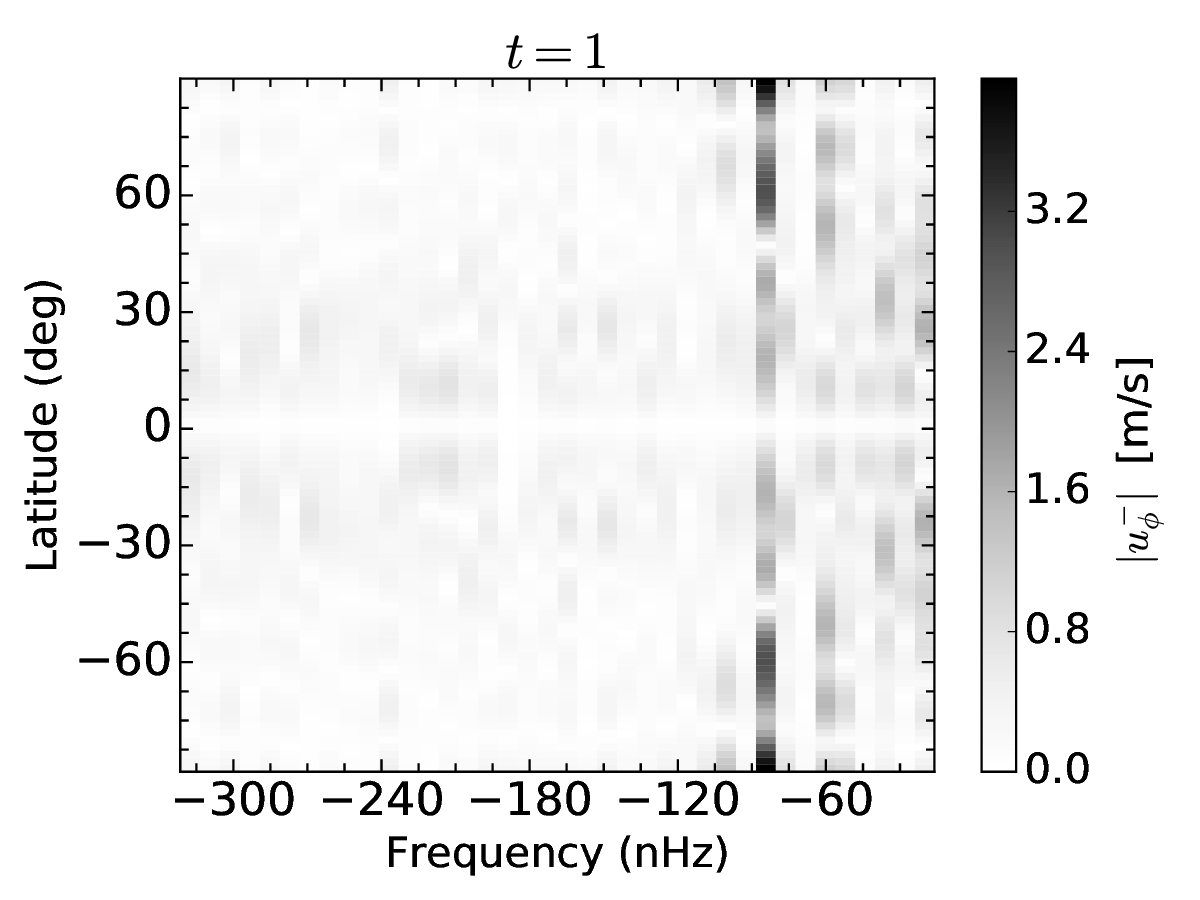}
  \includegraphics[scale=0.33]{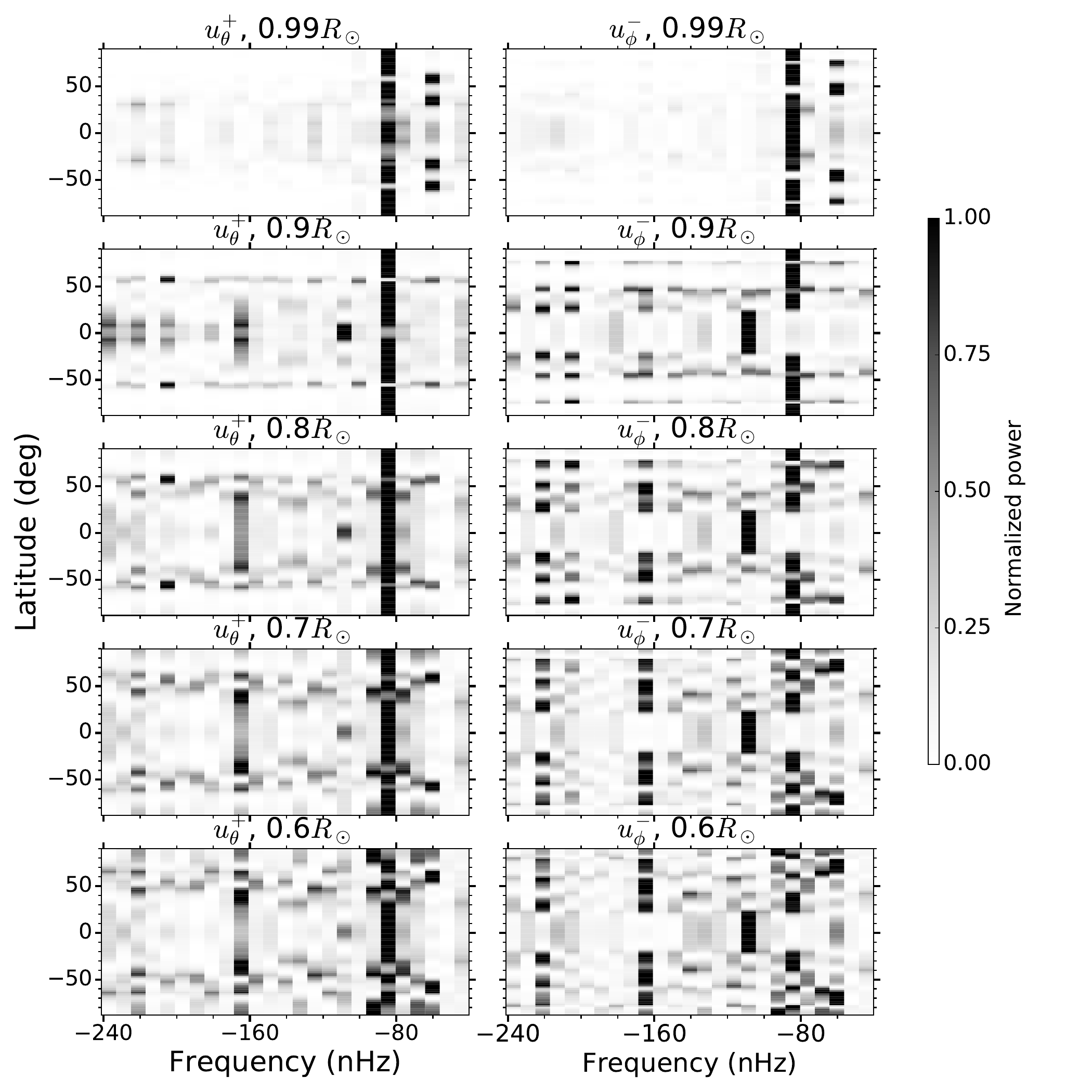} 
  \caption{Upper panels: power spectra of $u_{\theta}^{+}$ on the left and $u_{\phi}^{-}$ on the right. The mode amplitude exhibits significant strength at high latitudes, reaching magnitudes on the order of $4$ m/s. Bottom Panel: Depiction of normalized symmetric latitudinal velocity, $u_{\theta}^{+}$ (left panel), and antisymmetric azimuthal velocity, $u_{\phi}^{-}$ (right panel), corresponding to a symmetric mode with azimuthal order $t=1$. The depths at which inferences are made are specified in the titles of the respective panels. This illustration is based on a four-year dataset derived from SDO/HMI spanning from $2014$ to $2017$.}
  \label{fig:HL_depth}
\end{figure}

 The retrograde inertial mode with azimuthal order $t=1$, occurring at high latitudes (above $60^\circ$), exhibits the most pronounced amplitude, measuring approximately $10$ m/s. In order to investigate this prominent mode, we choose $\Delta\ell=0$ and an azimuthal order of $t=1$ to compute B-coefficients using Equation \ref{eq:BCoeff}. This choice facilitates the determination of $w^\sigma_{s,t=1}$ for all odd harmonic degrees, encompassing $s$ values ranging from $1$ to $20$. Subsequently, we employ Equation \ref{eq:uTor} to evaluate the hemispherically symmetric component $u_\theta$ and the antisymmetric component $u_\phi$. In line with the terminology outlined by \citet{gizon21}, these are referred to as $u_{\theta}^{+}$ and $u_{\phi}^{-}$, respectively. For the scope of this study, we direct our attention to the symmetric mode. To visualize the signal within the measured B-coefficients, we select the peak frequency $\sigma=-80.4$ nHz, where the signal is most prominent. Notably, these measured B-coefficients exhibit an alternation in signs relative to harmonic degrees. Consequently, we depict the values $(-1)^\ell B^\sigma_{st}$ in Figure \ref{fig:HL_bcoeff}. Importantly, the measured B-coefficient deviates from zero, displaying an increasing amplitude as $\ell$ grows. This trend implies the presence of a signal associated with the high-latitude mode. The amplification in intensity with higher values of $\ell$ is attributed to the $\ell^{3/2}$ factor present in the sensitivity kernel equation \citep[refer to Equation $8$ in][]{hanasoge18}. As long as the influence of the remaining terms in the sensitivity kernel remains subordinate, this increase in amplitude as $\ell$ rises is to be expected.
 We perform an inversion of the measured B-coefficients, yielding the radial profile $w^\sigma_{st}(r)$. Subsequently, we proceed to forward model the inverted $w^\sigma_{st}(r)$ and draw a comparison with the measured B-coefficients, as depicted in Figure \ref{fig:HL_bcoeff}. Figure \ref{fig:HL_sphere} showcases the representation of $u_{\phi}^{-}$ on a spherical surface, revealing a discernible spiraling pattern at higher latitudes. A similar spiral pattern was earlier reported by \citet{hathaway13,bogart15,hathaway2020,gizon21}. Additionally, we plot both $u_{\theta}^{+}$ and $u_{\phi}^{-}$ in Figure \ref{fig:HL_depth}, presenting them in a frequency versus latitude diagram. The power of this mode is indeed focused near the pole. Upon normalizing the power at each latitude, we observe a consistent presence of power around the frequency of $-80$ nHz across all latitudes, as depicted in Figure \ref{fig:HL_depth}. This widespread distribution suggests that this mode is global. Notably, the data used for the construction of Figure \ref{fig:HL_depth} spans a duration of four years, from $2014$ to $2017$. In an alternate assessment, we partition the $8$ years of SDO/HMI data, spanning from $2010$ to $2017$, into two separate $4$-year segments. In both scenarios, we observe that both $u_{\theta}^{+}$ and $u_{\phi}^{-}$ exhibit significant power centered around $80$ nHz. As anticipated, the signal from the lower depths appears noisier due to increased noise, yet we are able to measure it all the way down to the base of the convection zone.  To determine the central frequency and line-width of this mode, we create a power spectrum depicted in the lower panel of Figure \ref{fig:HL_depth}. We then apply a fitting procedure using a Lorentzian function augmented by a constant background term, defined as follows:
\begin{equation}
 F(\sigma)=\frac{A}{1+[(\sigma-\sigma_{0})/(\Gamma/2)]^2}+B,
 \label{eq:fit}
\end{equation}
In this equation, $A$ represents the amplitude, $\sigma_0$ denotes the central frequency, $\Gamma$ signifies the full width at half maximum, and $B$ stands for the background power. To execute the fitting process, we employ the {\it curve\_fit} module available in {\it scipy.optimize}. By applying this method to the first four years of SDO/HMI data (spanning $2010-2013$), we obtain a central frequency of $\sigma_0/(2\pi)=-80.1 \pm 2.1$ nHz and a line-width of $\Gamma/(2\pi)= 29.5 \pm 6.8$ nHz. Upon analyzing the data from SDO/HMI spanning the years $2014$ to $2017$, we determine the central frequency to be $\sigma_{0}/(2\pi)=-79.7 \pm 0.5$ nHz and the line-width as $\Gamma/(2\pi)= 7.4 \pm 2.0$ nHz. Subsequently, by averaging the power spectra from these two instances and performing fitting, we establish the central frequency as $\sigma_{0}/(2\pi)=-81.0 \pm 1.4$ nHz and the line-width as $\Gamma/(2\pi)=18.6\pm 4.4$ nHz. We depict the spatial representation of $u_\phi^{-}$ on a spherical surface in Figure \ref{fig:HL_sphere}, showcasing its depth variation. To assess the accuracy of our depth profile reconstruction, we subject it to a validation test against the differential rotation profile obtained through global helioseismology, as detailed in Appendix \ref{sec:diffRot}.
 
\subsection{Radial dependencies of Rossby modes} 
\begin{figure}   
      \includegraphics[scale=0.35]{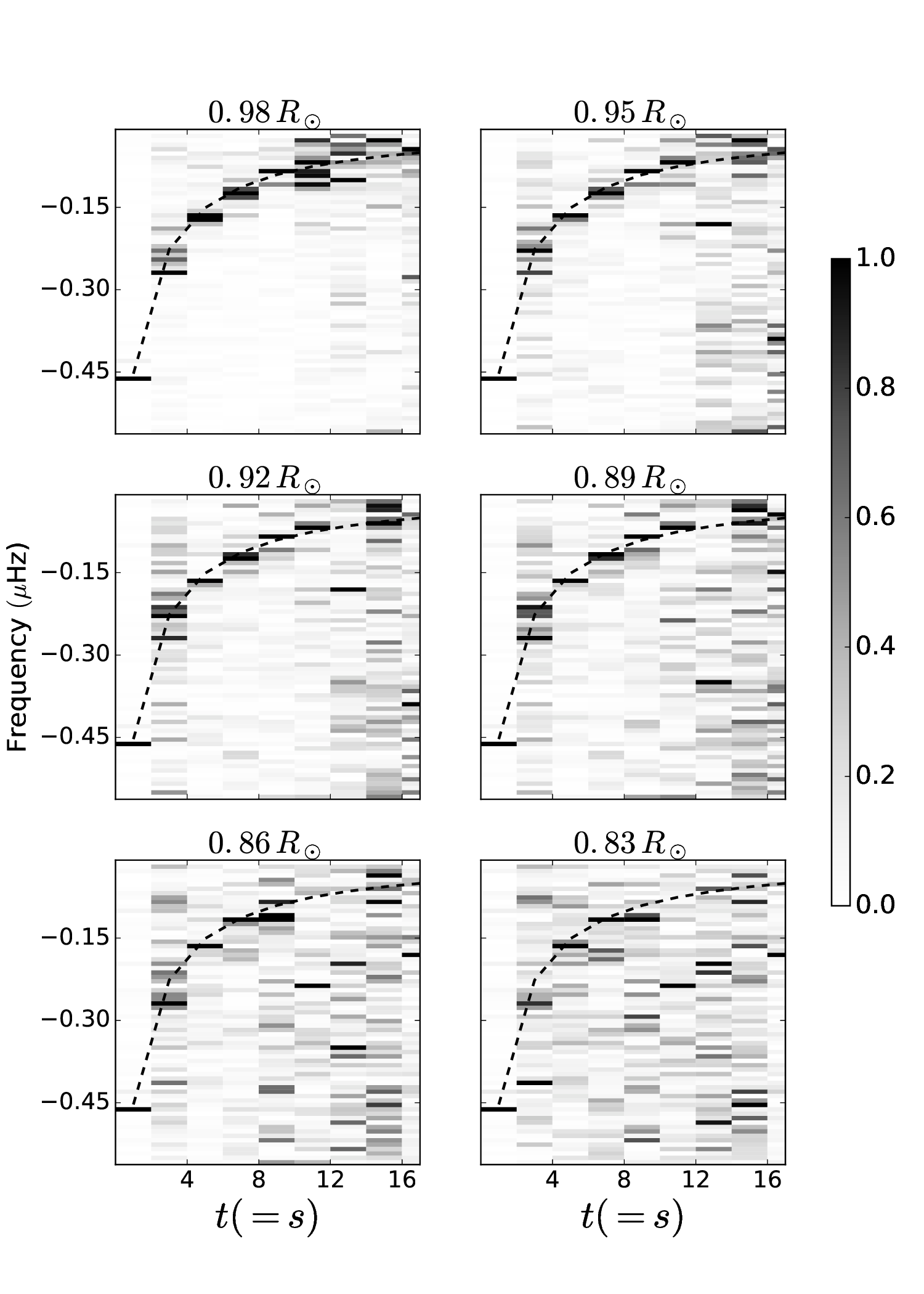}\includegraphics[scale=0.3]{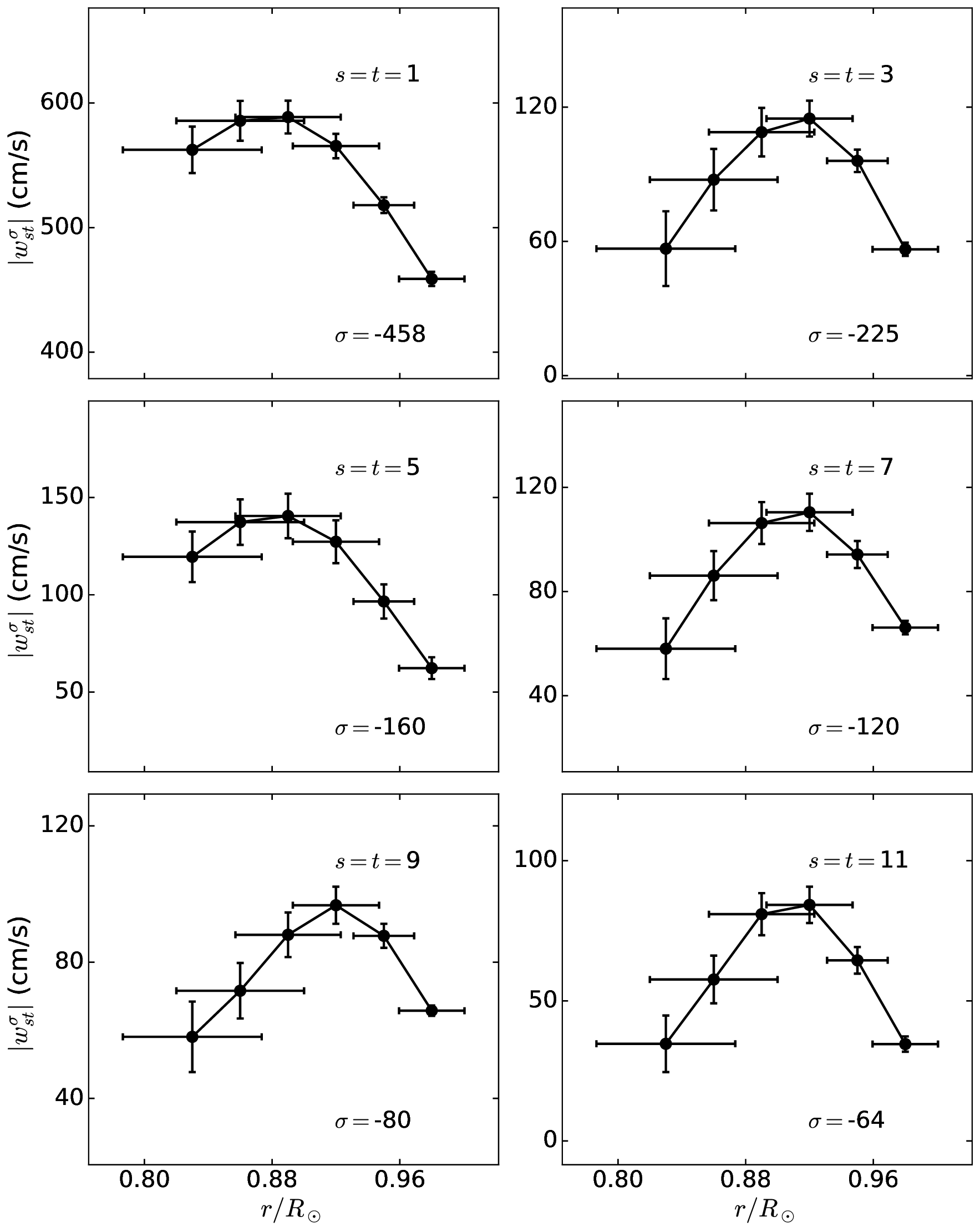}\\
      \includegraphics[scale=0.45]{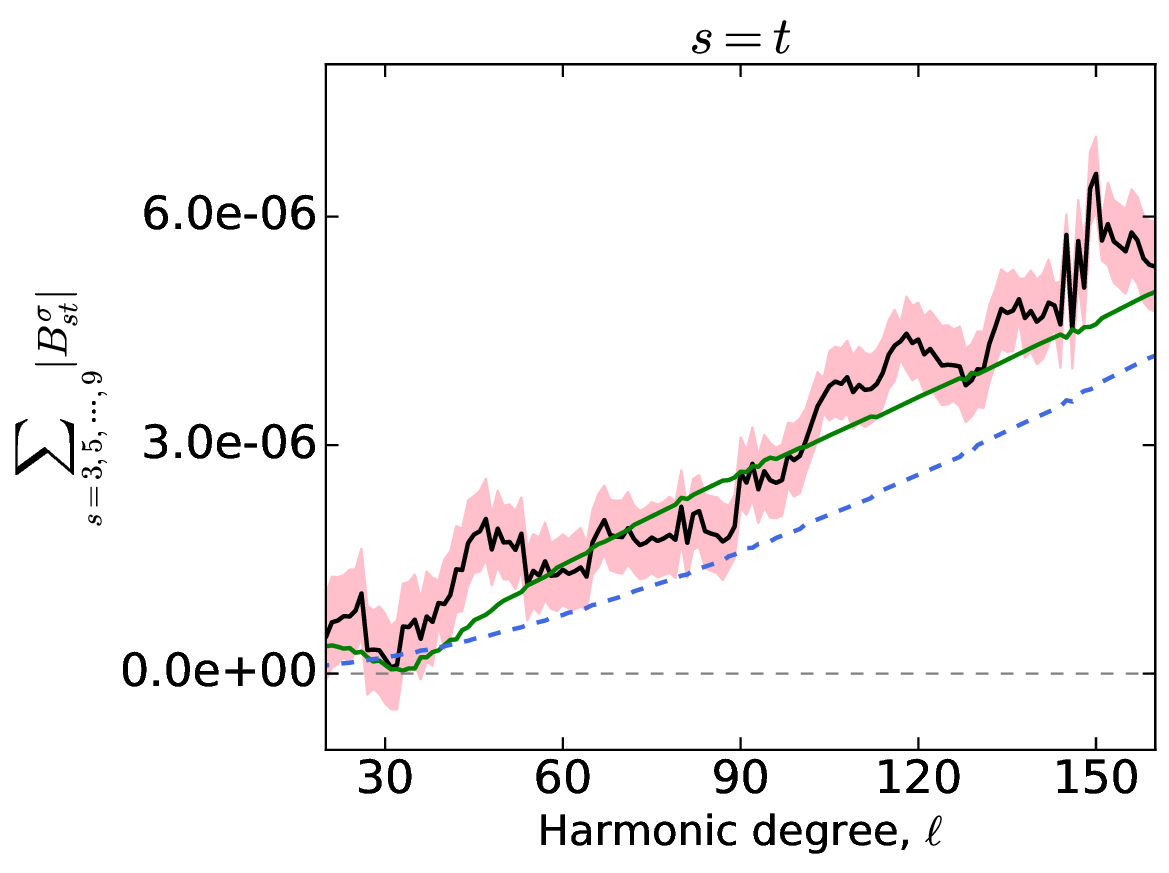}
    \caption{Upper left panel: normalized power $\vert w^\sigma_{st}\vert^2$ for odd harmonic degrees $s$ as obtained from inversion at various depths, each indicated in the panel header. Additionally, the black-dashed line in the graph represents the theoretical dispersion relation of Rossby waves in a uniformly rotating medium, with a rotation frequency of $\Omega/(2\pi)=453.1$ nHz. Upper right panel: absolute values of $w^\sigma_{st}$ are plotted as functions of radius for two distinct time periods, encompassing odd-harmonic degrees ranging from $s=1$ - $11$. The chosen frequency bin $\sigma$ corresponds to the maximum power for each mode, and its specific value is provided in each panel (measured in nHz). To construct these plots, a dataset spanning a total of eight years from 2010 to 2017, acquired from SDO/HMI, was divided into two four-year chunks. The inversion process made use of the average B-coefficients. Horizontal error bars in each panel represent the widths of the averaging kernels at the corresponding depths. Error bars are calculated based on two, four-year data series. Bottom panel: sum of measured B-coefficients for Rossby modes, encompassing all odd-harmonic degrees from $s=3$ to $s=9$. Frequency bins at which we find maximum power for each Rossby mode have been used in order to perform a similar analysis as in Figure \ref{fig:HL_bcoeff}. Reconstructed B-coefficients with the inverted profile are shown as the green solid line. If the depth profiles of Rossby modes had $r^s$ dependencies and maximum amplitudes is the same as observed at depth $0.99 R_\odot$, our estimated B-coefficients would correspond to the blue dashed line.
    }    
    \label{fig:dispersionDepth}
\end{figure}

Different values of $\Delta\ell$ can be used in Equation \ref{eq:BCoeff} to measure B-coefficients in order to image Rossby modes in depth \citepalias{mandal21}. We primarily consider $\Delta\ell=0$ case, which has the highest signal-to-noise. By inverting Equation \ref{eq:BInv} using B-coefficients where $s=t$, we generate the radial profile $w^\sigma_{st}(r)$. We illustrate the power $\vert w^\sigma_{st}\vert^2$ at various depths in Figure \ref{fig:dispersionDepth}. We expect significant power above the background close to the theoretical dispersion relation of Rossby waves, Equation \ref{eq:dispersionRossby}. As we go deeper, the background starts to dominate, and below $0.83R_\odot$, distinguishing signals for higher azimuthal orders, denoted as $t$, becomes challenging. However, traces of low azimuthal orders are still discernible in the plot; refer to Figure \ref{fig:dispersionDepth}. Our findings align with a similar conclusion when we investigate the weighted sum of measured B-coefficients using the phase-filter method described in Appendix \ref{sec:Rossby_phase}.  

 In order to determine how these mode amplitudes vary, we plot $w^\sigma_{st}$ as a function of radius for the frequency bins at which we observe the maximum power for each mode in Figure \ref{fig:dispersionDepth}. Rossby modes have significant power even in the deeper layers. We utilize a total of $8$ years of SDO/HMI data spanning $2010$ to $2017$, which we split into two distinct $4$-year segments. We consider average B-coefficients from these two data-sets and perform the inversion. We find that the amplitudes of all modes with odd harmonic degrees starting from $s=1$ to $s=11$ first increase with depth down to around $0.92 R_\odot$, decreasing at deeper layers. We perform similar experiments with the measured B-coefficients for Rossby modes as we did in Section \ref{sec:HL_mode} in order to test whether the observed B-coefficients deviate from the predicted B-coefficients when $w^\sigma_{st}$ assumes an $r^s$ variation. We first select frequency bins where the observed B-coefficient attains maximum power for each mode, and then add the signed B-coefficients so that they contribute positively in the bottom panel of Figure \ref{fig:dispersionDepth}. We compute theoretical B-coefficients for the synthetic $w^\sigma_{st}$ profile that varies as $r^s$. Bottom panel of Figure \ref{fig:dispersionDepth} shows that the synthetic B-coefficient deviates notably from the measured B-coefficients. 
    
 \section{Latitudinal eigenfunctions of Rossby modes}\label{sec:latEigen}
\begin{figure*}[h]  
 \centering
    \includegraphics[scale=0.35]{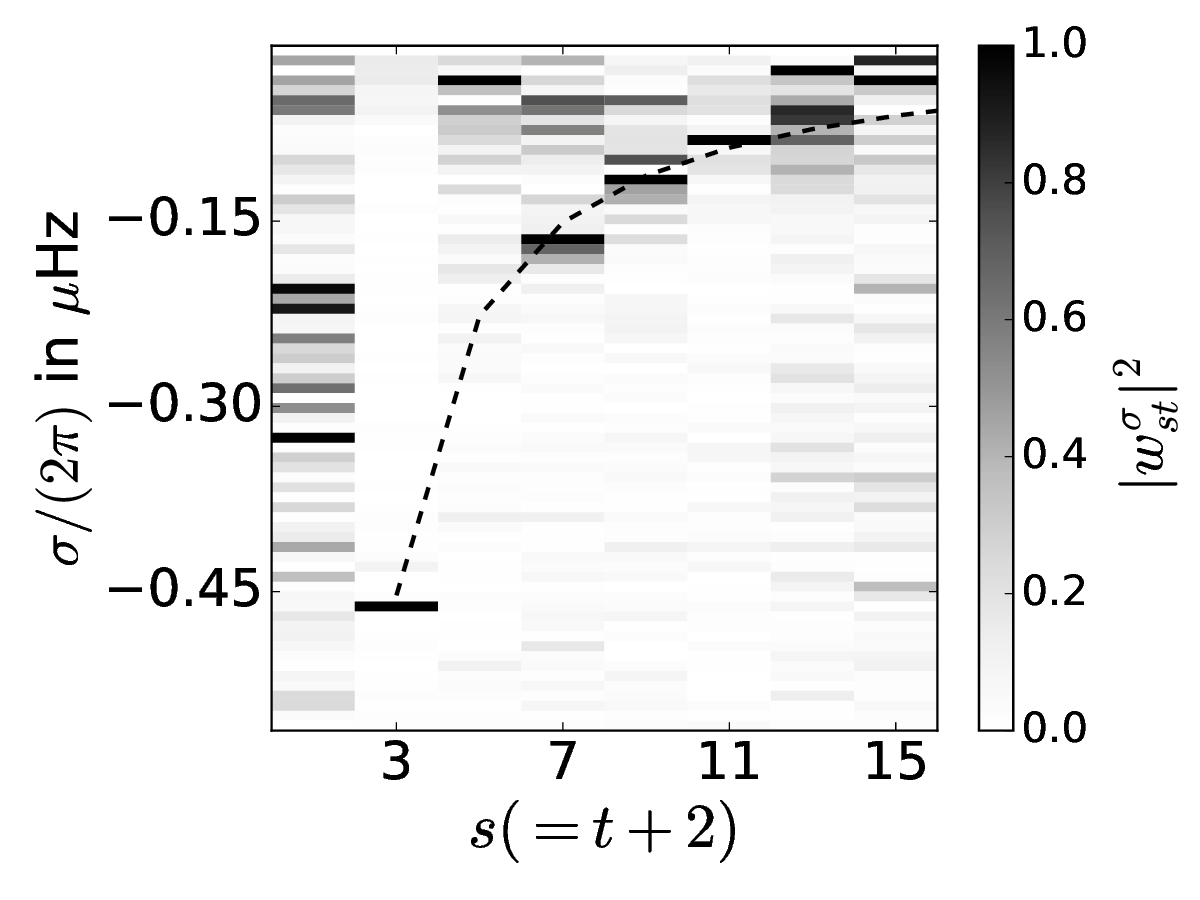} \includegraphics[scale=0.35]{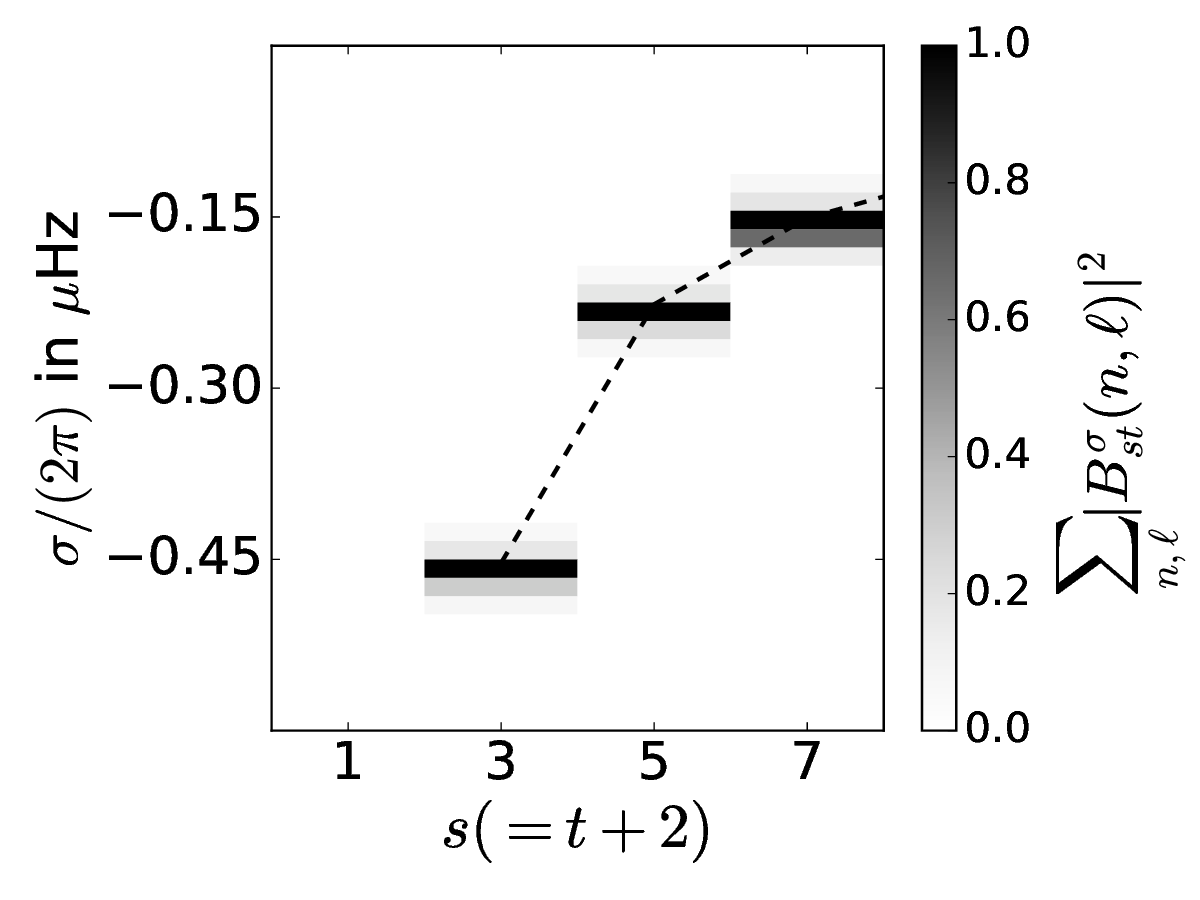}
    \includegraphics[scale=0.35]{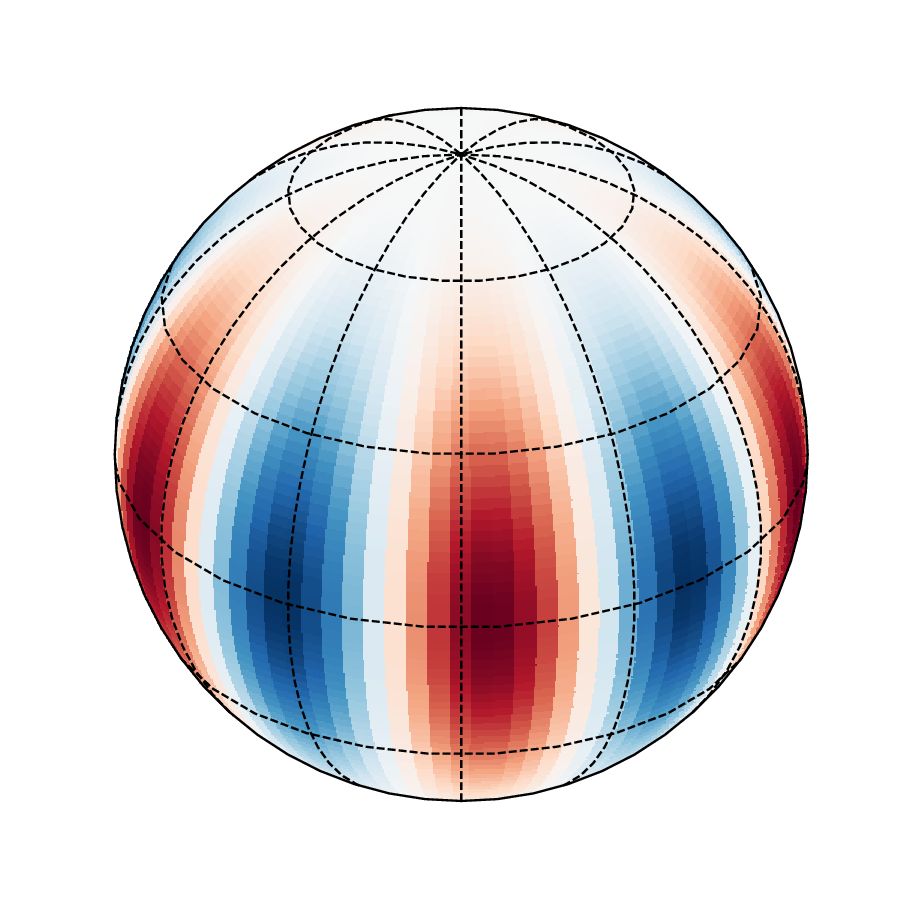}\includegraphics[scale=0.35]{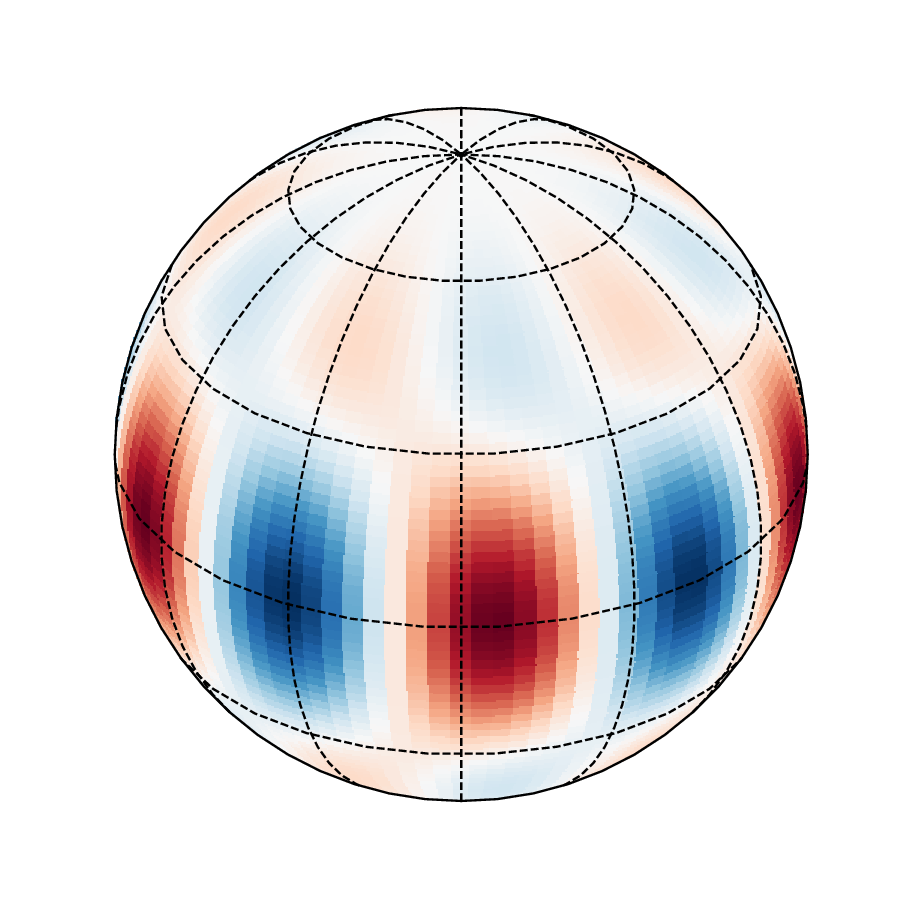}
     
    \caption{Upper left panel: Normalized power of $\vert w^{\sigma}_{st}\vert^2$ from inversions when $t=s-2$. $s$ varies from $1$ to $15$, assuming only odd values. Upper right panel: We observe leakage in our synthetic test, transitioning from mode $(s, s)$ to mode $(s+2, s)$. The black dashed line shows the function $-2\Omega/(s+1)$ for mode $(s+2,s)$ to highlight that leakage occurs at the same frequency as the mode, $(s,s)$. Lower left panel: Observed eigenfunction for the $t=5$ Rossby mode if we consider the mode to be sectoral. Lower right panel: Observed eigenfunction for Rossby mode if we consider contributions from different $s$ at fixed $t=5$. Bottom panel: Real part of $C_{t}(R,\theta)$ at all azimuthal orders, mentioned in each panel. Corresponding errors ($\pm 1\sigma$ around mean) in the analysis are denoted by the red shaded area.}
  \label{fig:latHMIOdd}
\end{figure*}

 Equation \ref{eq:uTor} implies that  latitudinal eigenfunctions are determined by derivatives of spherical harmonics, $\hat{\mathbf{r}}\times \nabla_{h}Y_{st}$.  In our prior studies \citepalias{hanasoge19,mandal21}, we have considered Rossby modes to be purely sectoral and only analyzed B-coefficients for $s=t$. Sectoral modes peak at the equator and decay towards higher latitudes but do not change sign. Whereas observational studies by \citet{gizon18,proxauf2020} found that surface eigenfunctions of these modes, which they label only with azimuthal number, are close to sectoral spherical harmonics but slightly different since they change sign at latitudes of $\sim\pm 30^\circ$. At fixed azimuthal order, the eigenfunctions contain contributions from both sectoral and non-sectoral components ($s\neq t$). The length scales of a perturbation are labeled by both harmonic degree, $s$ and azimuthal order, $t$ (see Equation \ref{eq:kernel_toroidal}) in normal-mode coupling and we analyze the measured B-coefficients for $s\neq t$. The latitudinal eigenfunctions of these modes deviate from sectoral spherical harmonics if contributions from the $s\neq t$ channels are sufficiently large. We analyze the B-coefficients for all possible cases where $s\neq t$, with the $s=\vert t\vert+2$ of particular interest. We find significant power as seen in Figure \ref{fig:latHMIOdd} and the observed power at the spatial scale $(s,t=s-2)$ follows the relation $-2\Omega/(t+1)$, which is also the frequency of the sectoral mode with azimuthal order $t$. In order to determine whether it is signal or spatial leakage from spatial scale $(s,s)$ to $(s+2,s)$, we perform similar synthetic tests as discussed in \citetalias{mandal2020}. We consider power only for sectoral modes $(s,s)$ and estimate B-coefficients that leak from sectoral to non-sectoral components, as shown in Figure \ref{fig:latHMIOdd}. It is seen that significant power from odd sectoral modes $(s,s)$ leak into spatial wave numbers $(s+2,s)$ at the same frequency bins. A similar  leakage phenomenon was also earlier reported by \citet{woodard21}. In appendix \ref{sec:appendix_a}, we explain why there would be a leakage similar to that observed in Figure \ref{fig:latHMIOdd}. 
 
\begin{figure*}          
     \includegraphics[scale=0.35]{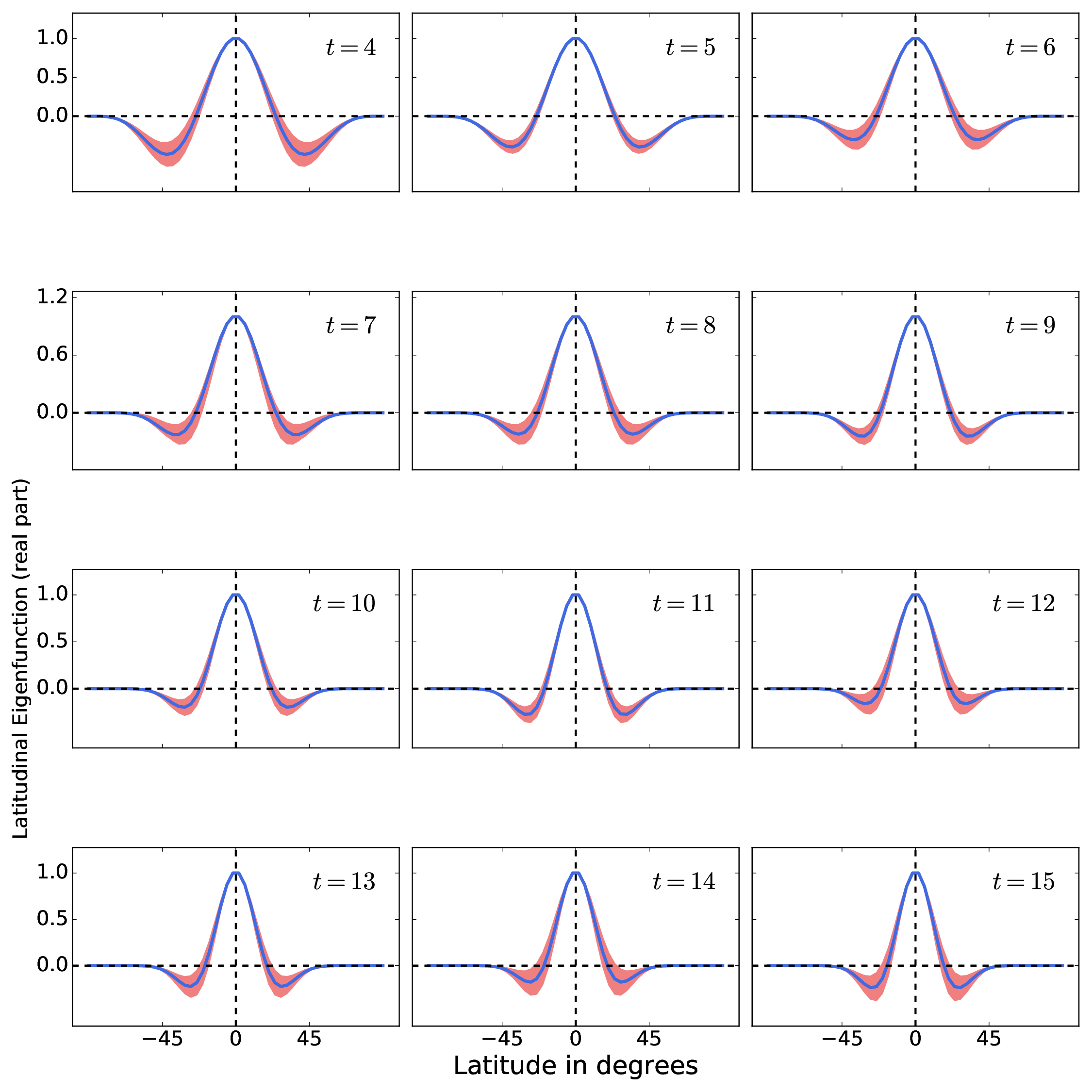}\includegraphics[scale=0.22]{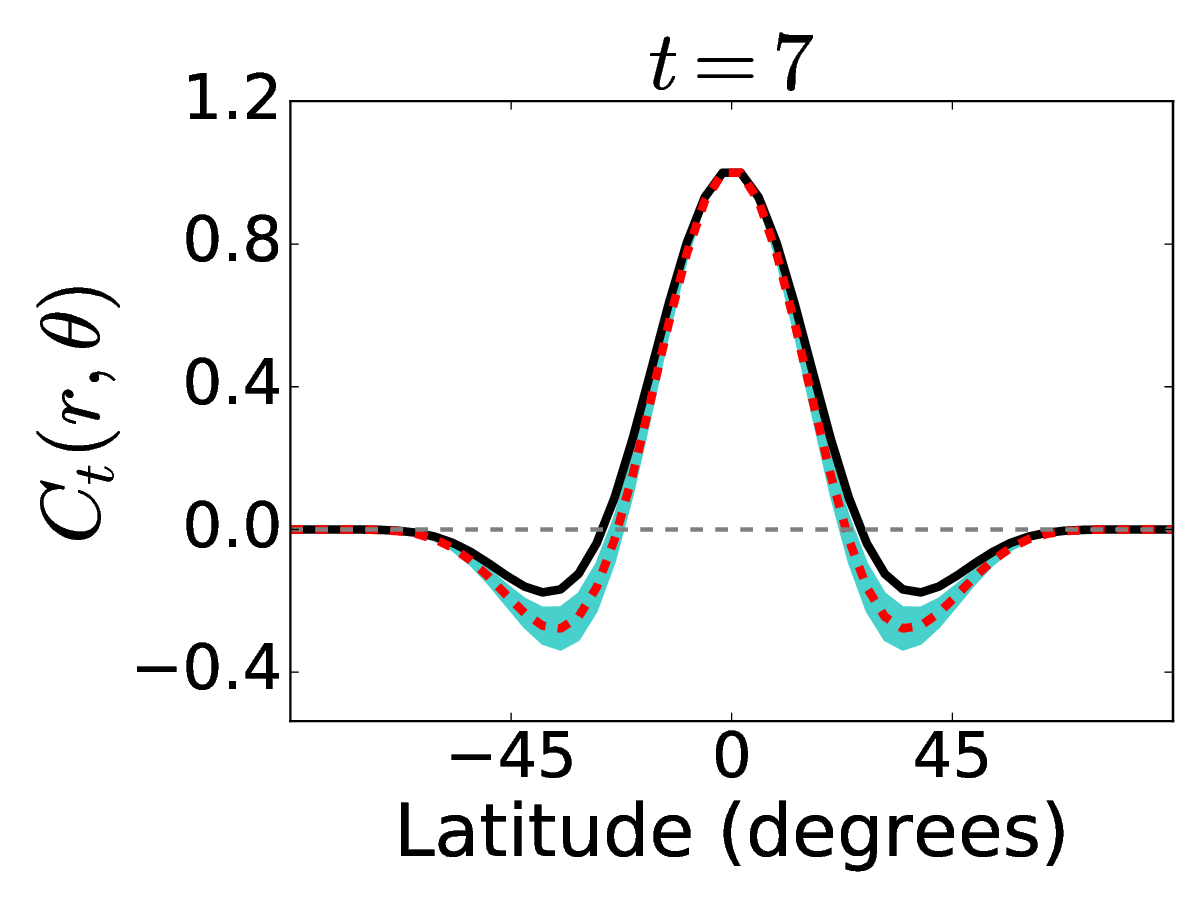}
     \caption{Left panel: Real part of $C_{t}(R,\theta)$ at all azimuthal orders, mentioned in each panel. Contributions from all $s\geq t$ have been taken into account. Corresponding errors ($\pm 1\sigma$ around mean) in the analysis are denoted by the red shaded area. Right panel: we use the inversion result for sectoral while non-sectoral components are represented by Gaussian noise. The noise level is varied to obtain $C_{t}(r,\theta)$. The black solid line is the real part of $C_t(r,\theta)$ when noise is comparatively low and the red-dashed line denotes the higher noise case. The shaded area corresponds to error estimates, $\pm 1\sigma$ around mean.}
     \label{fig:RossbyLatSphere}
 \end{figure*}

 We plot the latitudinal eigenfunction of the $t=5$ Rossby mode in Figure \ref{fig:latHMIOdd} for two cases. In one case, we consider it to be purely sectoral and in another, we consider contributions from the non-sectoral component $(s\neq t)$. It is seen that the sectoral-case eigenfunction does not have zero crossings at higher latitudes in the former scenario whereas it has zero crossings in the latter case.  Contamination by leakage as discussed above will make the interpretation of latitudinal eigenfunctions difficult in the present analysis. 
 Following the approach used by \citet{proxauf2020}, we estimate the temporal covariance of vorticity at each azimuthal order between the equator and all other latitudes according to
 \begin{equation}
C_t(r,\theta)=\frac{\langle \xi_t(T,r,\theta)\xi_t(T,r,\theta=90^\circ)\rangle_{t}}{\langle \vert\xi_t(T,r,\theta=90^\circ)\vert^2\rangle_{t}},
     \label{eq:Cov}
 \end{equation}
  where $\langle . \rangle_t$ denotes temporal averaging, $\xi_t$ is the centered vorticity $\xi^{\prime}_t=\xi_{t}-\langle \xi_t \rangle_{t}$ and $T$ denotes time. We show the estimated function $C(R,\theta)$ in Figure \ref{fig:RossbyLatSphere} for each $t$. We also find similar zero crossings at high latitudes as observed in previous studies and use a Monte-Carlo simulation to estimate errors in the inferences.\par
We consider another possibility, namely that $w^\sigma_{t+2\,t}$ is entirely noise. For this case, only contributions from sectoral components $w^\sigma_{tt}$ obtained from the inversion have been considered. We use a Gaussian noise model with zero mean and fixed standard deviation for $w^\sigma_{t+2\,t}$. We vary the noise amplitude and study its impact on the inference of the eigenfunction, $C_{t}$. We show our findings in Figure \ref{fig:RossbyLatSphere}. Zero crossings at high latitudes ($\sim \pm 30$) are also found in this case. We analytically explain this behavior in Appendix \ref{sec:appendix_b}. As the noise level increases, the absolute value of the minimum of $C_t$ also increases and the latitude at which it crosses zero approaches the equator. This study demonstrates that care must be taken to interpret the latitudinal eigenfunctions obtained using mode coupling, especially if we apply the definition given in Equation \ref{eq:Cov}. Our result shows that the presence of signal, leakage, and noise in non-sectoral components can all contribute to zero crossing. Hence, it is imperative to meticulously isolate each component and assess its individual contribution. This task warrants separate investigation, which we acknowledge as a crucial aspect for future studies. We underscored the importance of recognizing that leakage might be applicable to other methodologies as well, but the extent of its impact necessitates method-specific analysis.
 
 \section{Discussion and conclusions}
 Through our mode-coupling analysis, we identify inertial modes at high latitudes characterized by an azimuthal order of $t=1$ and a frequency of $\sim-80$ nHz. This substantiates the discovery of high-latitude characteristics initially reported by \citet{hathaway13,hathaway2020}, later confirmed as a normal mode by \citet{gizon21}. Our findings indicate the penetration of this mode throughout the entire convection zone. To explore dynamics in the polar region effectively, local helioseismic techniques necessitate accurate measurements at high latitudes. Unfortunately, current instruments do not have good coverage at the high latitudes. Mode coupling makes use of global p modes, which are generally very reliable measurements, and can be used in principle to detect critical and other inertial modes reported by \citet{gizon21}, which is part of set of future directions this work will take. We observe that our measured high latitude mode amplitude is approximately $4$ m/s, whereas \citet{gizon21}, \citet{hathaway13}, and \citet{bogart15} reported amplitudes on the order of $\sim 10$ m/s, $\sim 20$ m/s, and $\sim 0.5$ m/s, respectively. The task of measuring at high latitudes poses challenges owing to the restricted coverage offered by current instruments, a limitation initially noted in global helioseismology. Additionally this discrepancy may stem from the varying sensitivities of different methods to high latitudes, resulting in different amplitude measurements. Notably, this specific mode's amplitude peaks exclusively near very high latitudes. While this study presents an initial effort towards understanding high-latitude inertial modes with azimuthal order $m=1$, future work will address the challenges associated with measurements at high latitudes using normal-mode coupling. 
 \par
 Recently, \citet{Waidele23} established a correlation between the power and frequency of Rossby modes and the solar cycle, employing time-distance helioseismology and ring-diagram analysis. This correlation can also be investigated through mode coupling analysis, as demonstrated by \citet{hanasoge19}, who showed that even a two-year dataset from SDO/HMI can provide excellent signal-to-noise ratios for Rossby wave detection. Consequently, the mode-coupling technique emerges as a valuable tool for studying temporal variations in these waves and exploring potential correlations with solar cycle-related properties.     
 In earlier work by \citetalias{mandal2020,mandal21}, mode coupling was used to determine the eigenfrequencies and line-widths of Rossby modes. This work is aimed at investigating the depth structure of Rossby modes. We find that the amplitudes of these modes first increase with depth down to around $0.92 R_\odot$. To ensure the reliability of these inferences, we validate the technique by performing an inversion for the $s=3$ coefficient of differential rotation, as detailed in Appendix \ref{sec:diffRot}, which is seen to match with global helioseismology results \citep{larson18}. We do not intend to invert for spatial scales $s>3$ since they are significantly affected by leakage. A more thorough analysis to remove this effect needs to be designed for this purpose \citep[e.g.,][]{samarth21}.  We also add another validation tests in Appendix \ref{sec:rossby_test_app}. We choose a profile characterized by a radial node, with modes exhibiting amplitudes of $\sim 1$ m/s at the surface—a value in close proximity to the observed amplitude. In an effort to enhance the realism of the test, we perturbed the forward-modeled B-coefficients in alignment with the observed noise. We demonstrate that we are able to successfully recover the radial node, signifying that our analysis possesses the capability to reconstruct the input profile to a considerable extent. In  Figure \ref{fig:dispersionDepth} we find that Rossby modes can be observed for $r>0.83R_\odot$. With increasing inference depth, the background becomes stronger, and at $0.83 R_\odot$, it becomes impossible to discern these modes. We have used only $8$-years of SDO/HMI data to determine the depth dependence of Rossby waves. Analyzing extensive time series data is essential for enhancing the signal-to-noise ratio in deeper layers. Our initial endeavor involves inferring inertial modes in these regions, with plans for a comprehensive analysis utilizing all available data series in the future.  The error at the base of the convection zone is $\sim 15$ cm/s, calculated for a single frequency bin. Considering error propagation across other frequency bins, that contribute to the background, is crucial for detecting these modes in deeper layers. We observe that background noise becomes notably prominent as we approach $0.83R_\odot$. If  background noise remains low across these frequencies, it will allow us to probe deeper into the solar interior and identify signatures of modes in those deeper layers. Accounting for the potential temporal variability of Rossby waves, as demonstrated by \citet{Waidele23}, is crucial. Temporal averaging might therefore only provide an average profile for the specific period analyzed.
 \par
We also study the latitudinal eigenfunctions of these modes in Section \ref{sec:latEigen}. If modes are labeled only using azimuthal order, we expect contributions from both sectoral and non-sectoral components. We investigate the presence of signal in the non-sectoral components of these modes. We indeed find significant power in the $t=s-2$ channel, one source for which is power leakage from sectoral components $(s,s)$ to the non-sectoral components $(s+2,s)$. In a more comprehensive analysis, we will need to model and remove these leakage contributions in order to determine the latitudinal eigenfunction. Without mitigating these systematics, we estimate the latitudinal eigenfunctions (Left panel of Figure \ref{fig:RossbyLatSphere}), inferring that they have a zero crossing at high latitudes around $\pm 30^\circ$. We also investigate how the conclusion will be affected if there were no leakage from sectoral to non-sectoral components but the non-sectoral component were purely noise. We consider Gaussian noise for non-sectoral components and we set the sectoral component to the inversion result. In this case also, we find similar zero crossings as shown in  the right panel of Figure \ref{fig:RossbyLatSphere}. This study demonstrates that care must be taken to interpret the latitudinal eigenfunctions obtained using mode coupling, especially if we apply the definition given in Equation \ref{eq:Cov}. 
 The amplitude of the Rossby mode is approximately $1$ m/s, rendering their imaging at depth a challenging endeavor. Nevertheless, the high latitude inertial waves with azimuthal order $t=1$ possess a greater strength compared to Rossby modes, facilitating a more manageable analysis. These high-latitude inertial modes extend throughout the entire convection zone. Looking ahead, we hold an optimistic outlook that by broadening the spectrum of couplings and refining the measurement methodology, we can achieve an enhanced signal-to-noise ratio. This, in turn, will enable us to image the deeper layers with increased accuracy.  
 \par
\section*{}
\begin{flushleft}
\textit{Acknowledgments}: We express our gratitude to the anonymous referee for his/her valuable comments, which have contributed in improving the manuscript. K. M. appreciates the support from the ERC Synergy Grant WHOLESUN 810218 during his time at MPS, as well as the support from NASA grants 80NSSC20K0602 and 80NSSC20K1320 while at NJIT. Also, K. M. acknowledges helpful discussions with Alexander Kosovichev, Yuto Bekki, Jesper Schou, Laurent Gizon, and Aaron Birch. 
 \end{flushleft}
 \setlength{\parskip}{0pt}
 
 \appendix
  \section{Validating mode coupling by inferring solar differential rotation} \label{sec:diffRot}
\begin{figure}    
 \centering
     \includegraphics[scale=0.45]{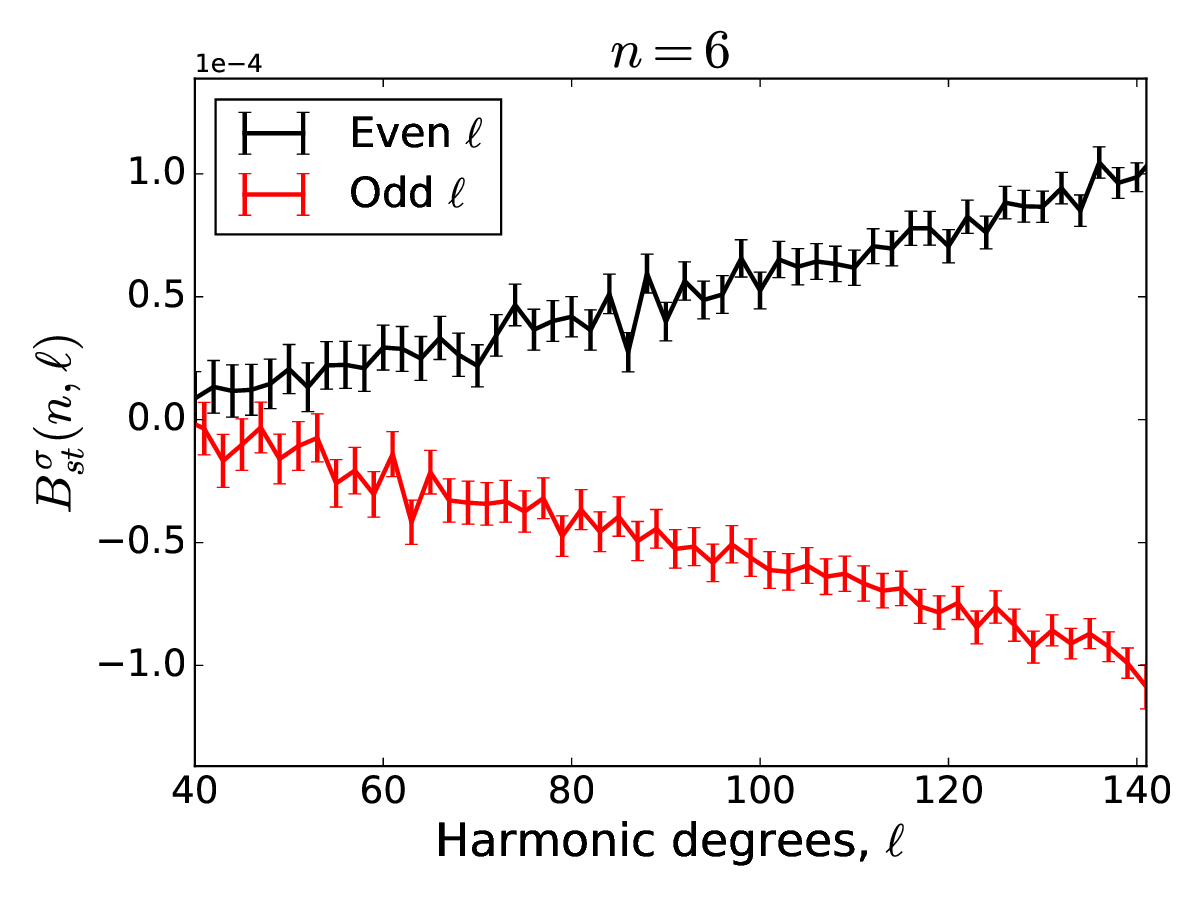}\includegraphics[scale=0.45]{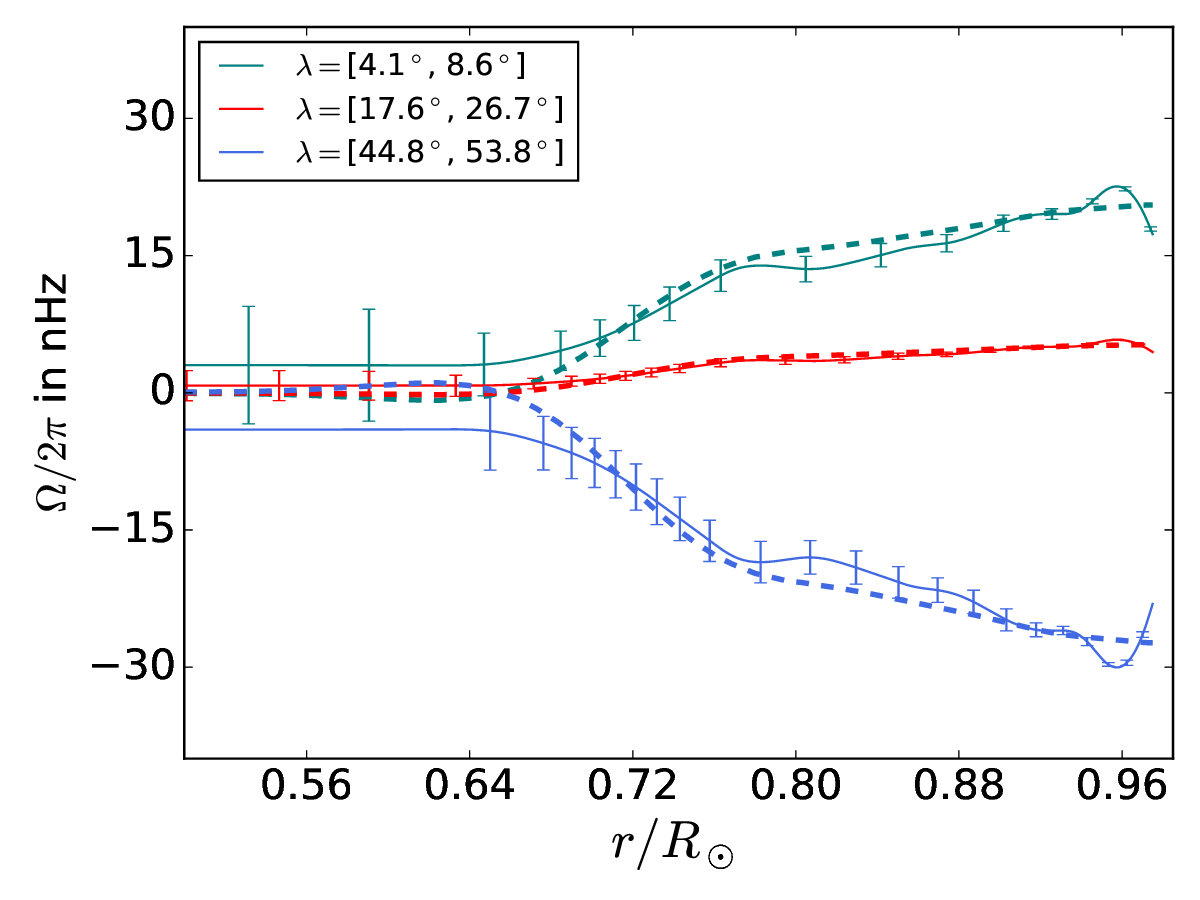}     
     \caption{Left panel: We show measured B-coefficient values from one year of SDO/HMI data with odd and even harmonic degrees, $\ell$ separately. It is seen that the signs for these two cases are opposite to those indicated by the expression for sensitivity kernel. Right panel: We compare inversions for $w_{s=3;t=0}\frac{\partial Y_{s=3;t=0}}{\partial \theta}$ from mode coupling (solid lines) with the same spatial component from global helioseismology (dashed lines) for three different latitude ranges. The latitude range over which averaging is performed is stated in the panel.}
     \label{fig:bcoeffOddEven}
 \end{figure}

 Prior studies by \citet{woodard13, schad2020, samarth21,das_srijan} have harnessed mode coupling to deduce solar differential rotation. In this work, we strive to achieve this inference through a distinct approach. We adhere to the measurement methodology previously employed by \citetalias{mandal21}. Since rotation is a steady axisymmetric flow, we need to measure the B-coefficient at $\sigma=0$ and $t=0$ in Equation \ref{eq:BCoeff}.  We consider $\Delta\ell=2$ and $t=0$ to measure the coupling induced by rotation in Equation \ref{eq:BCoeff}, using one year of SDO/HMI data. Because of the selection rule mentioned subsequent to Equation \ref{eq:fS}, we cannot estimate rotation at spatial scale $s=1$ for the mode-coupling measurement with $\Delta\ell=2$. The $s=1$ component of differential rotation becomes measurable when we analyze self-coupling among acoustic modes, which essentially constitutes the power spectrum of acoustic modes. The influence of the $s=1$ component of rotation is limited to the first order of the power spectrum. Since the power spectrum is more dominant for that case, we address leakage from power spectrum by employing the leakage matrix $L_{\ell m}^{\ell^\prime m^\prime}$, akin to the approach used in global helioseismology. The next significant components for rotation are $s=3,5$. Spatial leakage may be an important systematical effect to account for; however, we do not attempt to model this in the present work. The analytical formulation for the sensitivity kernel, denoted by $\mathcal{K}_{n\ell}(r)$, incorporates a prefactor of $(-1)^{\ell} \ell^{1.5}$, as outlined in Equation $8$ of \citet{hanasoge18}. This expression suggests an anticipated pattern in the B-coefficients we measure, displaying alternating signs in correspondence with the harmonic degree, $\ell$. This phenomenon is illustrated in Figure \ref{fig:bcoeffOddEven}. Specifically, B-coefficients for odd harmonic degrees exhibit a sign contrary to those associated with even harmonic degrees. This distinct sign pattern attests to the efficacy of our measured B-coefficients in capturing the underlying signal. Consequently, we proceed to perform an inversion to derive $w^\sigma_{st}$, which we subsequently compare with the rotation profile obtained through global helioseismology \citep{larson18}. We show our results in Figure \ref{fig:bcoeffOddEven}.  
Despite using only one set of couplings and only one year's worth of data, it is possible to estimate $w^\sigma_{st}$ for $s=3$ from the surface down to the base of the convection zone. While we observe minor discrepancies, particularly in the vicinity of the surface beyond $0.97 R_\odot$, the underlying reasons for these deviations remain unclear, necessitating further investigation in future studies. We do not attempt to estimate $w^\sigma_{st}$ for $s=5$ as we expect leakage from the more dominant $s=3$ component. Therefore we need to take into account leakage if we want to analyze the $s=5$ component, which we reserve for future work. The data analysis procedure used in previous studies with mode coupling are different from the data reduction technique applied here. This validation exercise gives us confidence in our attempt to infer radial eigenfunctions of inertial modes in the Sun. 

 \section{Rossby modes in the interior layers via B-coefficients}\label{sec:Rossby_phase}
\begin{figure}   
\centering
    \includegraphics[scale=0.4]{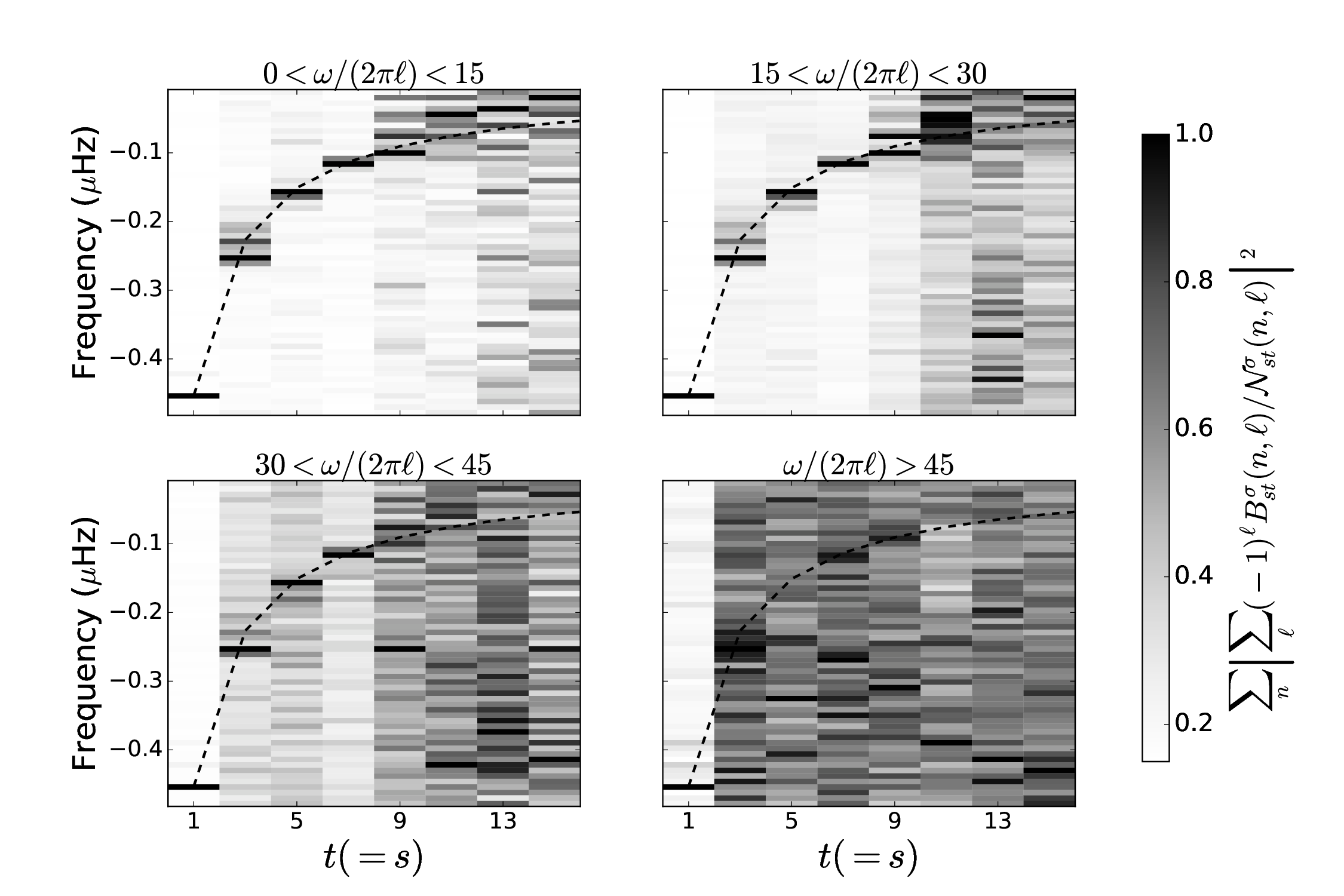}
    \caption{We apply phase-speed filters and estimate $\sum_{n}\vert\sum \frac{(-1)^\ell B^\sigma_{st}(n,\ell)}{\noise^{\sigma}_{st}(n,\ell)}\vert^2$ using the observed B-coefficients. The applied filters are described in the panel titles. Here, $\nu$ is in $\mu$Hz. In the bottom-right panel, where the applied filter is $\nu/\ell>45$, power close to the theoretical dispersion relation does not stand out from the background, in contrast to other panels.}
    \label{fig:bcoeffPhase}
\end{figure}

 We estimate the sum of the signed B-coefficients, $(-1)^\ell B^\sigma_{st}(n,\ell,\ell+\Delta\ell)$ over all $\ell$ used in the analysis and compute the sum of the squares of their absolute values over all radial orders, $n$, in order to estimate the quantity $\sum_{n}\vert\sum_{\ell} (-1)^\ell B^\sigma_{st}(n,\ell,\ell+\Delta\ell)/\noise^{\sigma}_{st}(n,\ell,\ell+\Delta\ell)\vert^2$, where $\noise^{\sigma}_{st}(n,\ell,\ell+\Delta\ell)$ corresponds to measurement noise. This quantity is only a function of two variables, $\sigma$ and $t$. We apply a phase-speed filter, as these describe the regions through which acoustic modes traverse the interior. We have, $\omega/\ell=c(r_t)/r_t$, where $r_t$ corresponds to the inner turning points of the p-modes and $c$ is the sound speed at that depth. We thus indirectly infer how deep the Rossby waves penetrate into the interior. We show our results in Figure \ref{fig:bcoeffPhase}, which characterizes different limits on $\omega/\ell$. We find that, when $\omega/\ell>45$, it becomes difficult to discern signal from background ($\omega/\ell=45$ corresponds to depth $r_t=0.83R_\odot$.)

 \section{\label{sec:appendix_a} Effect of leakage in determining latitudinal eigenfunctions of Rossby modes}
   The observed line-of-sight Doppler velocity cannot be exactly decomposed into spherical harmonics due to our inability to observe the far side of the Sun. This results in a smearing in spectral space termed ``leakage", resulting in cross-mode talk, i.e., $(\ell^\prime,m^\prime)$ leaks into the mode of interest, $(\ell,m)$, given by 
  \begin{equation}
      \Phi_{\ell m}=\sum_{\ell^\prime m^\prime}L_{\ell m}^{\ell^\prime m^\prime} a_{\ell^\prime m^\prime},
  \end{equation}
  where $L_{\ell m}^{\ell^\prime m^\prime}$ quantifies the geometric leakage from mode $(\ell,m)$ to ${\ell^\prime,m^\prime}$.
  Because of this limitation, our measured B-coefficients, $B^\sigma_{st}$, with harmonic degree, $s$ and azimuthal order, $t$, are contaminated by contributions from other spatial scales $s^\prime,t^\prime$, mathematically expressed as
  \begin{equation} B^\sigma_{st}=\sum_{s^\prime,t^\prime}\mathcal{L}_{st}^{s^\prime t^\prime}b^\sigma_{s^\prime t^\prime},\label{eq:A_Bst_b}
  \end{equation}
  where $b^\sigma_{s^\prime t^\prime}$ is the true B-coefficient in absence of systematics, $\mathcal{L}_{st}^{s^\prime t^\prime}$ denotes leakage from spatial scale $(s^\prime,t^\prime)$ to $(s,t)$, which may be expressed in terms of the leakage matrix, $L_{\ell m}^{\ell^\prime m^\prime}$ \citepalias[see for][]{mandal2020}.
 If $\sigma_{t;i}$ and $\sigma_{t;c}$ are the frequencies of sectoral Rossby mode measured in the inertial and co-rotating frames respectively, they are related through
  \begin{align}
      \sigma_{t;i}=-\frac{2\Omega}{s+1}+t\Omega,\\
      \sigma_{t;i}=\sigma_{t;c}+t\Omega.
  \end{align}
 Rewriting the frequency term in Equation \ref{eq:A_Bst_b} as measured from the inertial frame,
 \begin{equation}
    B^{\sigma_{t;c}+t\Omega}_{st} =\sum_{s^\prime t^\prime}\mathcal{L}_{st}^{s^\prime t^\prime}b^{\sigma_{t;c}+t\Omega}_{s^\prime t^\prime},\label{eq:BSt}
 \end{equation}
  where $b^\sigma_{st}$ is significant only when $\sigma_{t;c}$ is close to the sectoral-Rossby-mode frequency  $\sigma_t=-2\Omega/(s+1)$. For simplicity, we describe it using a delta distribution,
  \begin{equation}
      b^\sigma_{s^\prime t^\prime}=\delta_{s^\prime t^\prime} \delta[\sigma-(\sigma_{t;i})].\label{eq:bSt}
  \end{equation}
 Substituting Equation \ref{eq:bSt} into Equation \ref{eq:BSt}, we obtain
 \begin{equation}
     B^{\sigma_{t;c}+t\Omega}_{st}=\sum_{s^\prime} \mathcal{L}_{st}^{s^\prime s^\prime} \delta[\sigma-(\sigma_{s^\prime;i})].
 \end{equation}
 For $t=s-2$ and frequency $\sigma=\sigma_{s-2,i}$,
 \begin{equation}
     B^{\sigma_{s-2;c}+(s-2)\Omega}_{s\; s-2}=\sum_{s^\prime} \mathcal{L}_{s\;s-2}^{s^\prime s^\prime} \delta[\sigma_{s-2}+(s-2)\Omega-(\sigma_{s^\prime;i})].
 \end{equation}
 The right hand side is only significant when $s^\prime=s-2$.  
 \begin{equation}
    B^{\sigma_{s-2;c}+(s-2)\Omega}_{s\; s-2}=  \mathcal{L}_{s\:s-2}^{s-2\: s-2} \delta[0].\label{eq:A_B0}
 \end{equation}
 Equation \ref{eq:A_B0} tells us that we can expect leakage from mode $(s-2,s-2)$ at spatial scale $(s,s-2)$ because of the leakage matrix $\mathcal{L}_{s\,s-2}^{s-2\,s-2}$ at frequency $\sigma_{s-2;c}$ in the co-rotating frame, which is also the frequency of the sectoral mode, $(s-2,s-2)$, explaining our findings in Figure \ref{fig:latHMIOdd}.   
 
 \subsection{\label{sec:appendix_b} Measurement of vorticity}
  Vorticity may be calculated by using the velocity function defined in the equation below,
  \begin{align}
      \xi_{t}(T)=\sum_{s,t^\prime}\int_{-\infty}^{\infty} d\sigma e^{i\sigma T}s(s+1)w^\sigma_{st}(R)\int_{0}^{2\pi} d\phi Y_{st^\prime}(\theta,\phi)e^{-it\phi}\nonumber\\
      =2\pi\sum_{s}\sqrt{\frac{(2s+1)(s-t)!}{4\pi(s+t)!}}\int_{-\infty}^{\infty} d\sigma e^{i\sigma T}s(s+1)w^\sigma_{st}(R)P_{st}(\theta),
      \label{eq:vor}
  \end{align}
  where $P_{st}$ is the associated Legendre polynomial of harmonic degree $s$ and azimuthal order $t$. We drop the summation over $s$ in Equation \ref{eq:vor} and only consider contributions from two terms for simplicity.  
  \begin{equation}
      \xi_t=\chi_{tt}(T,r)P_{t\,t}(\cos\theta)+ k \chi_{t+2\,t}(T,r)P_{t+2\,t}(\cos\theta), 
      \label{eq:chi}
  \end{equation}
  where $k$ is a constant, $\chi_{st}$ is the vorticity. For sectoral Rossby modes, $\chi_{t+2\,t}$ can contribute to $\xi_{t}$ as a leak from $\chi_{t\,t}$ or as pure noise in the measurement. We consider these two cases separately below.
  \subsection{\label{sec:AP_l}First case: Leakage}
  If $\chi_{t+2\,t}$ is due to leakage from $\chi_{t\,t}$, the above Equation \ref{eq:chi} reduces to
  \begin{equation}
      \xi_{t}=\chi_{tt}(T,r)(P_{t\,t}(\cos\theta)+\mathcal{L}_{t\,t}^{t+2\,t}P_{t+2\,t}(\cos\theta)),
      \label{eq:xi}
  \end{equation}
  where $\mathcal{L}_{tt}^{t+2\,t}$ denotes leakage from mode $(t,t)$ to $(t+2,t)$. Substituting Equation \ref{eq:xi} into Equation \ref{eq:Cov}, we obtain
  \begin{equation}
      C_{t}(r,\theta)=\frac{\langle\vert \xi_{tt}(T,r)\vert^2\rangle(P_{tt}(\cos\theta)+\mathcal{L}_{tt}^{t+2\,t}P_{t+2\,t}(\cos\theta))}{\langle \vert\xi_t(T,r,\theta=90^\circ)\vert^2\rangle}.
      \label{eq:Ct}
  \end{equation}
   We expect a zero crossing from Equation \ref{eq:Ct} because of the term $P_{t+2\,t}$.
   
  \subsection{Second case: only noise, no leakage}
  We model $\xi_{t+2\,t}$ as  uncontaminated by leakage (discussed in section \ref{sec:AP_l}) but rather, as pure noise. Therefore, we rewrite this term
  \begin{equation}
      \xi_{t+2\,t}=N(0,\mu)(t,r)P_{t+2\,t}(\cos\theta),
      \label{eq:xi_n}
  \end{equation}
  where $N(0,\mu)$ is zero-mean Gaussian noise with standard deviation $\mu$. Substituting Equation \ref{eq:xi_n} into \ref{eq:Ct}, we obtain
  \begin{equation}
      C_t(r,\theta)=\frac{\langle\vert \xi_{tt}(T,r)\vert^2\rangle P_{tt}(\cos\theta)+\vert N(0,\mu)\vert^2 P_{t+2\,t}(\cos\theta)}{\langle \vert\xi_t(T,r,\theta=90^\circ)\vert^2\rangle}.
      \label{eq:Cn}
  \end{equation}
 We see from Equation \ref{eq:Cn} that, even if $\xi_{t+2\;t}$ were to be pure noise, leaked contributions from it could result in zero crossings of latitudinal eigenfunctions, as observed in right panel of Figure \ref{fig:RossbyLatSphere}. 
 
 \section{Synthetic tests for Rossby  waves profile} \label{sec:rossby_test_app}
\begin{figure} 
    \includegraphics[scale=0.42]{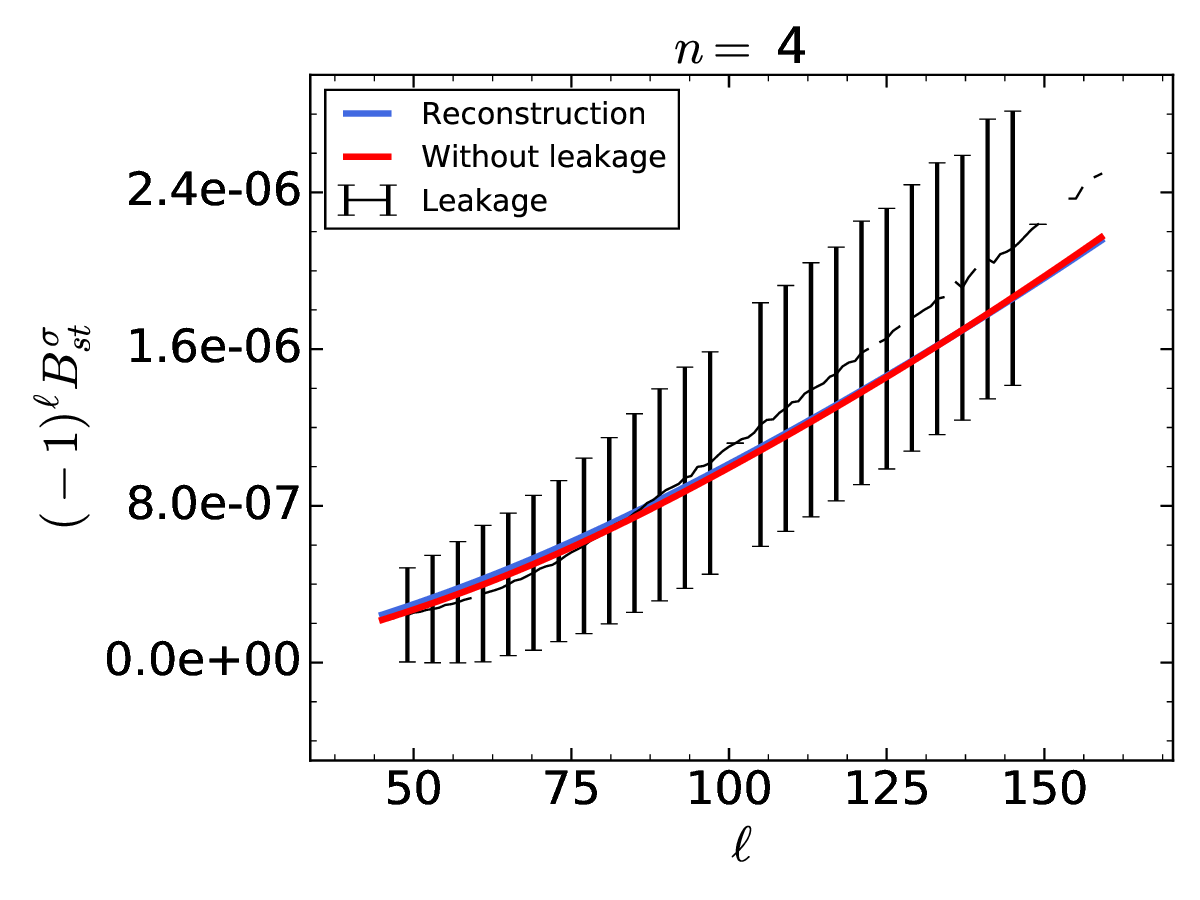}\includegraphics[scale=0.42]{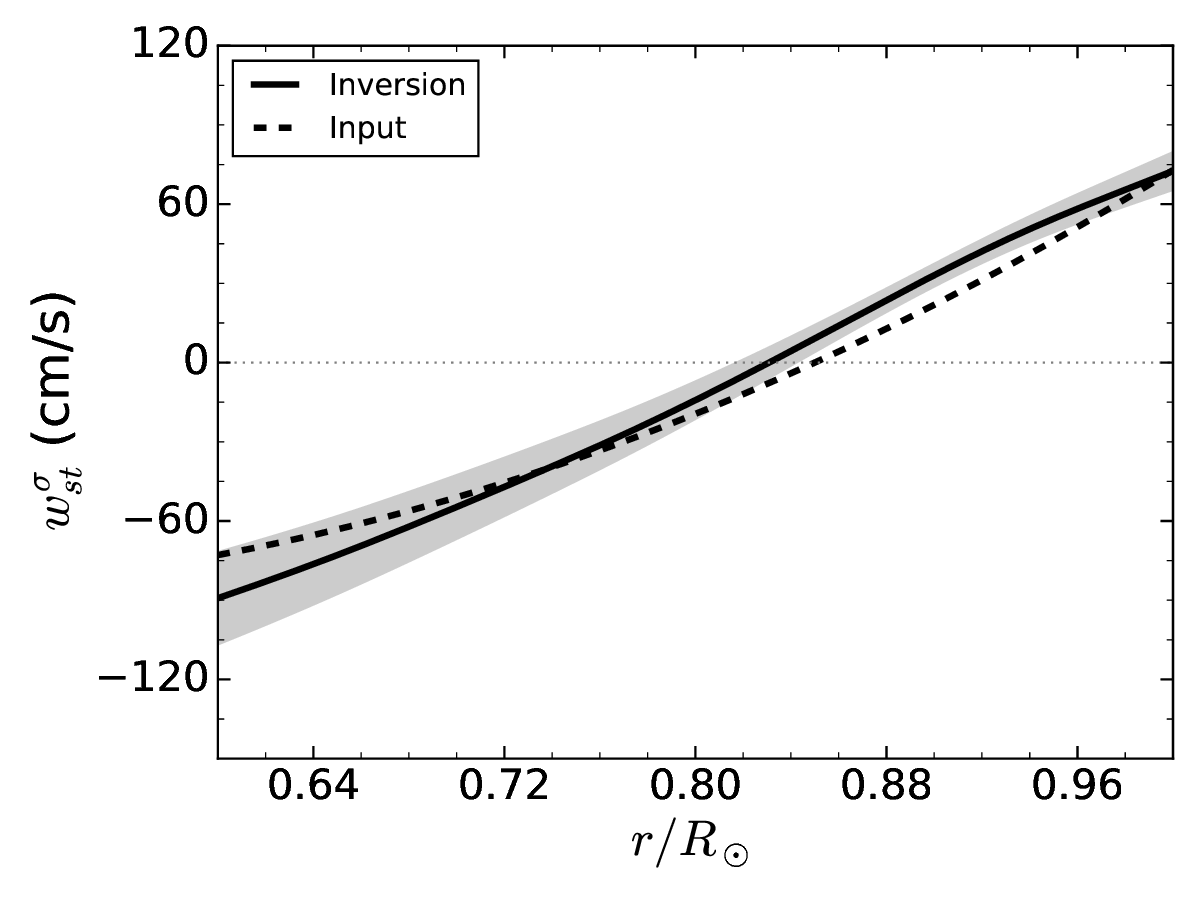}
    \caption{Left panel: we plot the signed B-coefficient, $B^\sigma_{st}$ (depicted as a black solid line with error bars), for harmonic degree $s=3$ and $t=3$, and for radial order $n=4$ (as specified in the panel title), corresponding to the frequency bin where the (s=3, t=3) Rossby modes peak. Corresponding $b^\sigma_{st}$ which is not affected by leakage is shown by red solid line. We perform forward modeling using the inverted profile of $w^\sigma_{st}$ based on Equation \ref{eq:B_apprx}. The resulting reconstructed B-coefficient is represented by the blue solid line. Right panel: Inverted profile of $w^\sigma_{st}$ with harmonic degree, $s=3$ is shown by black solid line. Corresponding error in the inverted profile ($\pm 1\sigma$) is denoted by gray shaded area.}
    \label{fig:syn_test}
\end{figure}

  In this study, we select a Rossby wave profile characterized by nodes positioned at a depth of $0.85 R_\odot$. Our objective is to explore the possibility of recovering nodes along the radial direction. This investigation is motivated by theoretical analyses \citep{bekki22_linear, jishnu22} indicating that Rossby modes may exhibit either zero or one radial node. We generate B-coefficients, $B^\sigma_{st}$ which is related to true $b^\sigma_{st}$ using following  equation \citealp[Equation $8$ of][]{mandal2020}

 \begin{equation}
     B^\sigma_{s, -s}(n,\ell)=N^\sigma_{\ell st}\sum_{\ell^\prime,\ell^{\prime\prime},m,m^\prime,s^\prime
      }L_{\ell m}^{\ell^\prime,m^{\prime}}L_{\ell m-s}^{\ell^{\prime\prime}m^\prime-s^\prime}\gamma_{-s \,m}^{\ell s\ell}H^{\sigma *}_{\ell\ell m\,-s}
      \gamma^{\ell^{\prime\prime}s^\prime\ell^\prime}_{-s m^\prime}H^{\sigma}_{\ell^{\prime}\ell^{\prime\prime}m^{\prime}-s^\prime}b^\sigma_{s^\prime ,-s^\prime}(\ell^\prime,\ell^{\prime\prime}).\label{eq:B_leak}
 \end{equation}
where $b^\sigma_{st}$ is 
\begin{equation}
    b^\sigma_{st}(n,\ell)=f_{\Delta\ell,s}\int_{0}^{R_\odot} dr w_{st}^{\sigma}(r)\mathcal{K}_{n\ell}(r)),
    \label{eq:b_true}
\end{equation}
We select a radial profile for $w^\sigma_{st}$ of Rossby modes and employ Equation \ref{eq:b_true} to estimate $b^\sigma_{st}$. In the absence of any leakage (i.e., when $L_{\ell m}^{\ell^\prime m^\prime}=\delta_{\ell \ell^\prime}\delta_{m m^\prime}$), it becomes evident from Equation \ref{eq:B_leak} that $B^\sigma_{st}$ simplifies to $b^\sigma_{st}$.To assess the impact of leakage on the modification of $B^\sigma_{st}$, we adopt a profile for $w^\sigma_{st}$ given by $A r(r-0.85) f(\sigma,\sigma_{s})$, where $f(\sigma,\sigma_{s})$ represents a Lorentzian profile with a central frequency $\sigma_s$ derived from the dispersion relation of Rossby waves. The parameter $A$ is selected to ensure that the amplitude of the modes closely approximates $1$ m/s at the surface, aligning with observed values. Following forward modeling with this $w^\sigma_{st}$ profile to derive $b^\sigma_{st}$, we substitute it into Equation \ref{eq:B_leak} to determine $B^\sigma_{st}$.  Our tests reveal that when $s^\prime \neq s$, there is no contribution since the modes have distinct frequencies, $\sigma$. Instead, contribution arises from neighboring modes, $(\ell^\prime, m^\prime)$. The left panel of Figure \ref{fig:syn_test} illustrates the disparity between $B^\sigma_{st}$ and $b^\sigma_{st}$. The figure highlights that the impact of leakage becomes more noticeable for higher $\ell$, yet the changes in values are not overly significant, which bodes well for the analysis. In order to check whether we can recover the original profile, we invert the following Equation 
\begin{equation}
   B^\sigma_{st}(n,\ell)= f_{\Delta\ell,s}\int_{0}^{R_\odot} dr w_{st}^{\sigma}(r)\mathcal{K}_{n\ell}(r)).
   \label{eq:B_apprx}
\end{equation}
 Equation \ref{eq:B_apprx} mirrors Equation \ref{eq:b_true}, with the only difference being the substitution of the left-hand side from $b^\sigma_{st}$ to $B^\sigma_{st}$. This substitution serves as an approximation that we employ in our analysis. To make it more realistic we perturb $B^\sigma_{st}$ according to noise from observation. We present our inverted profile in Figure \ref{fig:syn_test} and compare it with the original profile. We successfully recover the original profile with a moderate degree of accuracy, including the nodal point radius. To quantify the uncertainties in the inverted profile, we perturb $B^\sigma_{st}$ according to observed noise, generating $100$ realizations. For each realization, we perform inversion to obtain $w^\sigma_{st}$ and calculate the variance in the radial profile. This process allows us to determine error bars in the inverted profile of $w^\sigma_{st}$. Based on our analysis, we conclude that the approximation $B^\sigma_{st}\approx b^\sigma_{st}$ is reasonably accurate. In the future, we will explore avenues for refining this approximation to achieve a more precise profile.  
  
\bibliography{arxiv_submission} 
  
\end{document}